\documentclass[12pt,letterpaper]{article}
\usepackage[a4paper, total={7in, 10in}]{geometry}

\usepackage{graphicx}
\usepackage[utf8]{inputenc}
\usepackage{indentfirst}
\usepackage[english]{babel}
\usepackage{empheq}
\usepackage{stackengine}
\usepackage{helvet}
\usepackage{authblk}
\usepackage{hyperref}
\usepackage{comment}
\usepackage[noabbrev,capitalise,nameinlink]{cleveref}
\usepackage{amsmath} 
\usepackage{amssymb} 
\usepackage{orcidlink} 
\usepackage[super,comma,sort&compress]  
   {natbib}
\usepackage[right]{lineno} \linenumbers

\usepackage{mathtools}

\makeatletter
\renewcommand{\maketitle}{\bgroup\setlength{\parindent}{0pt}
\begin{flushleft}
  \textbf{\@title}
  
  \@author
\end{flushleft}\egroup}
\makeatother

\title{Reinforcement Learning for Control of Evolutionary and Ecological Processes}
\date{}

\author[1,2,3,*]{Bryce Allen Bagley}
\author[1,2]{Navin Khoshnan}
\author[1,2,4,5]{Claudia K Petritsch}

\affil[1]{Mathematical Medicine Group, Stanford University, Stanford,
CA 94305, USA}
\affil[2]{Department of Neurosurgery, Stanford Medical School, Stanford,
CA 94305, USA}
\affil[3]{Physician-Scientist Training Program, Stanford Medical School, Stanford,
CA 94305, USA}
\affil[4]{Maternal \& Child Health Research Institute, Stanford University School of Medicine, Stanford,
CA 94305, USA}
\affil[5]{Stanford Cancer Institute, Stanford University School of Medicine, Stanford, CA 94305, USA}

\affil[*]{Correspondence: bbagley@stanford.edu}

\begin{document}

\maketitle

\section*{SUMMARY}

As research in evolutionary dynamics and Evolutionary Game Theory moves from the realm of theory into application in oncology, algorithms that account for physiologic variation are needed to move beyond simple models. Yet few algorithms exist that integrate these ecological and physiological factors which are central to evolution in biological systems. We introduce a new model of evolutionary games which accounts for ecology and physiology by framing both as computations. Leveraging this new model, we develop an algorithmic approach for controlling evolution via methods from Reinforcement Learning, a branch of Machine Learning, and prove rigorous algorithmic bounds on the complexity of solving this problem. This approach enables us to obtain first-of-their-kind results on the algorithmic problem of learning to control an evolving population of cells. We prove a complexity bound on controlling evolution in situations with limited prior knowledge of cellular physiology or ecology, give the first results on the most general version of the mathematical problem of directed evolution, and establish a new link between AI and biology.

\section*{KEYWORDS}


reinforcement learning, evolution, evolutionary dynamics, artificial intelligence, evolutionary game theory, quantitative oncology, mathematical oncology

\section*{INTRODUCTION}

In medical contexts, the clearest application of \textit{Evolutionary Game Theory} (EGT).\cite{nowak_evo_dynamics,rice_evolutionary_theory,basanta_cancer_ecology,sandholm_population_games_evolutionary_dynamics,cotner_2023_integrating_quantitative_methods_and_math_for_modeling_cancer_dynamics,wolf_2021_evolutionary_game_theory_contribution_understanding_treating_cancer,gatenby_cancer_therapy_evo_dynamics} is to modeling the physiology of cancers and their responses to therapies with the goal to optimize cancer treatment.\cite{Pucci_cancer_treatment_current_perspectives_2019,Nikolaou_drug_resistance_cancer_treatment_challenges_2018,gatenby_cancer_therapy_evo_dynamics,Tonorezos2024,Delgado-López_glioblastoma_review_2016} There is growing interest within clinical medicine regarding the possibility of applying EGT to the development and optimization of cancer therapies, immunology, and infectious disease treatment, but thus far there has been limited ability to integrate the enormous quantities of available biomedical data. In addition, there are very few studies of treatment courses with meaningful dynamics over time, particularly because of the difficulty of experimenting with combination and time-course therapies. Instead, the current standard of care in cancer is generally to use one or two chemotherapeutics along with surgery and/or radiation. If the patient is not cured, the cancer develops resistance to those drugs through multi-faceted mechanisms, and a new drug or drug combination must be used. Mechanisms of resistance include, among others, modulation of the immune system, changes to non-coding regions which interfere with therapy efficacy, and mutations and epigenetic changes which alter features of cancer cells such as signalling pathways, drug transport into and out of cells, target site structure, and drug metabolism.\cite{Tufail2024} However, as discussed in \cite{gatenby_cancer_therapy_evo_dynamics, basanta_cancer_ecology,wolf_2021_evolutionary_game_theory_contribution_understanding_treating_cancer,cotner_2023_integrating_quantitative_methods_and_math_for_modeling_cancer_dynamics} and many others, there is growing evidence that therapies which vary over time in more strategic and complex ways have great potential to improve cancer outcomes. Similarly, bacterial and viral infections develop resistance to antibiotics and antivirals via evolutionary processes very much analogous to those in cancers. 

EGT uses combined elements of the discrete games from economics, the differential games primarily from engineering, and the mathematical and biological literature on evolution and ecology to predict how evolutionary processes determine the frequencies of different types in a population.\footnote{The term "types" in EGT is a deliberately general one, as the mathematical tools can be used for the study of genotype frequencies in a population, foraging strategies in macro-organisms, cultural changes, and even collective behavior in response to economic policies.} In biomedical contexts, there has been particular interest in the application of EGT to the study of cancer and therapy resistance,\cite{basanta_cancer_ecology,cotner_2023_integrating_quantitative_methods_and_math_for_modeling_cancer_dynamics,wolf_2021_evolutionary_game_theory_contribution_understanding_treating_cancer,gatenby_cancer_therapy_evo_dynamics,mcevoy_cancer_game_theory_limitations,stankov_optimizing_cancer_therapy_egt,orlando_cancer_egt_optimal_control_chemo} and this is likewise our focus in this paper. 

A key goal of predicting evolutionary dynamics is to be able to intervene and steer evolution towards a desired state. This would represent a quantitative form of directed evolution, an approach of artificial selection already used on bacteria to optimize biomolecular catalysts for industrial manufacturing of pharmacologics, agricultural chemicals and more.\cite{packer_directed_evolution_proteins,kumar_directed_evolution_biocatalysts,wang_directed_evolution_method_apps_review,porter_directed_evolution_industrial_catalysts} This is essentially an approach of applying \textit{artificial} selection in much the same fashion that \textit{natural} selection optimizes organisms in nature, and the existing experimental success is indicative of the possibilities which may lie in more quantitatively-driven use. Accomplishing this will require mathematical tools for guiding dynamics in a fine-grained manner, and such tools will themselves require the development of new theoretical and computational results. 

Here we explore two key ideas. First, questions of how to prove a rigorous algorithmic bound on the computational problem of optimal control of evolution. Second, the duality in the nature of the underlying process as both a competition and a computation. The former has clear connections with what we discuss above about the history and present state of the cross-talk between game theory and evolutionary biology.\footnote{This conceptual view is ubiquitous in theoretical biophysics, systems biology, and adjacent fields, and a recent review article offers a helpful explanation of them in terms of potential therapeutic applications.\cite{lagasse_molecular_collective_intelligence}} The latter, however, is perhaps a subtler one, and is related to a radical yet compelling proposal by Krakauer, Bertschinger, Olbrich, Flack, and Ay.\cite{krakauer_2020_information_theory_of_individuality} They propose that individuality exists not as a binary property, but a spectral one with its origins in information theory. Instead of arbitrary human-imposed boundaries, one instead looks at how the mutual information of different partitionings varies and thus defines degrees of individuality. This will be discussed further in the Methods section on Computational Duality.

Our goal in this paper is to develop a novel approach which introduces new techniques into the theoretical literature of EGT while focusing on its ultimate use in experiment and application. To accomplish this, we make use of Osband and Van Roy's work on model-based reinforcement learning (RL) of Markov Decision Processes, in which they proved the first unified regret bounds for model-based RL of quadratic systems -- like EGT -- which to our knowledge remain state of the art to this day.\cite{osband_van_roy_eluder_dim_model_based_reinforcement_learning,osband__russo_van_roy_posterior_sampling_reinforcement_learning} Linking these results from the theory of artificial intelligence with the specific function classes\footnote{A term used in machine learning and artificial intelligence to denote a collection of functions with one or more relevant properties in common.} of interest in EGT, we introduce and develop a mathematical formalism for an evolutionary control theory via a combination of EGT, reinforcement learning, and control systems. In addition to defining and exploring this formalism, we derive bounds on learning to control directed evolution which match the state-of-the-art bounds in the theory of artificial intelligence more broadly. However, much of the value of theory comes from eventual application to practice, and as discussed by Traulsen and Glynatsi there is a growing divide between theory and practice in EGT.\cite{Traulsen_2023_the_future_of_theoretical_evolutionary_game_theory} 

In the standard formulation of EGT,\cite{nowak_evo_dynamics,rice_evolutionary_theory,sandholm_population_games_evolutionary_dynamics} one studies a vector $c$ of $q$ types -- $c_1, c_2, ..., c_q$ -- which interact within some environment with the goal of predicting what fraction of the total population each type will comprise over time. The interactions between types are modeled as an interaction matrix $\mathbf{F}$ such that $f_{ij}$ predicts how the $j$-th type will impact the $i$-th type. The unadjusted fitness of a type $c_i$ -- that is, the rate at which its population will grow -- is thus defined by $\mathbf{F}_ic $, where $\mathbf{F}_i$ is the $i$-th row of $\mathbf{F}$. Because we are concerned with what fractions of the population each type comprises, these fitnesses must be normalized. Thus, the average fitness of all types is computed via weighting by population fraction, and this is subtracted from the fitnesses of all types to normalize population growth. 

While in reality populations essentially never remain static in total size, this fractional representation is far more useful for mathematical analysis and in general does not diminish the value of the analysis. The growth of a type's sub-population is proportional to its current size, so the total formulation of a basic Evolutionary Game Theory problem is as follows.

For a specific type, 

\[
    d_t c_i = c_i \left(\left(\sum_j \mathbf{F}_{ij} c_j \right) - c^{\hspace{1mm}T}\hspace{1mm}\mathbf{F}c\right)
\]

Extending to all types present, we have

\begin{equation}\label{default_EGT_equation}
    d_t c = \mathbf{C}_a \left(\mathbf{F}c - \vec{\hspace{0.5mm}1}c^{\hspace{1mm}T}\hspace{1mm}\mathbf{F}c\right)
\end{equation}

Here $\mathbf{C}_a$ is a diagonal matrix whose entries are the elements of $c$, defined explicitly as
\begin{equation}\label{C_matrix_definition}
    \mathbf{C}_a = \sum_{i=1}^q \mathbf{E}_i c^T\vec{e}_i
\end{equation}

 Where here $\vec{e}_i$ is the vector with all zero entries save for a $1$ at entry $i$, and $\mathbf{E}_i$ is the $q\times q$ matrix with the $q$-th main diagonal entry as 1 and all other entries as 0. We have subscripted this matrix $\mathbf{C}_a$ for convenience and cleaner notation in certain more complex calculations to come.

Additionally, $c^{\hspace{1mm}T}\hspace{1mm}\mathbf{F}c$ is the scalar representing the population-fraction weighted sum of raw fitnesses of all types. This can be more understood as first computing the fitness of each type and then taking the weighted average for the population by multiplying each individual fitness by its population fraction and summing.

This general form of nonlinear dynamical system can thus be used to predict how the composition of the overall population will change over time. Extensions often account for mutations, which in the continuous case essentially transfer infinitesimal fractions of the population from one type to another, but the general principle holds.\cite{nowak_evo_dynamics,rice_evolutionary_theory,sandholm_population_games_evolutionary_dynamics,may_ecology}

If mutations are to be incorporated, this would be done via the addition of a linear mapping $\mathbf{M}$ which describes the average rate at which each respective type mutates into any others. This mapping would be a symmetric matrix with diagonal entries $1-\delta$, with the off-diagonal entries representing these mutations being some distribution which sums to $\delta$.

Compared with the deeply interlinked nature of theory and data analysis within bioscience in the first half of the 20th century, comparatively little interaction occurred between them in the latter half. Despite being theoretical, EGT has still provided useful insights even in abstract yet practical contexts. It offers explanations of why intuitively strange situations in ecology can remain stable, and in recent years has seen early application to analyzing cancer progression and microbial evolution of antibiotic resistance. However, these efforts have in some ways been constrained by the methods used. In most applications the number of types $q$ might be three or four, and the entries of $\mathbf{F}$ will be bird's-eye-view estimates from large-scale observations of bacterial and cancer cell populations.

Bioinformatics tools which have emerged in recent decades may allow quantitative researchers to generate higher-fidelity approximations of $\mathbf{F}$ for $q$ large enough to make clinically relevant predictions beyond the simple test cases being assessed at present. Such work is critical, and a goal of this study is to begin the creation of a new class of tools for such researchers to use. In particular, work in the area of network inference algorithms has made it increasingly possible to estimate interaction effects within cells and between them. This represents an opportunity for a new approach to evolutionary dynamics which can make use both of contemporary methods in biomedical data sciences \textit{and} of rich theory.

\section*{RESULTS}\label{section_results}

Modeling the biophysics of disease physiology across multiple scales and diverse types of transport and biochemical processes is a significant mathematical challenge. As discussed in the Methods subsection on Computational Duality, one of the new methods we introduce is to model multiple types of biophysical processes as all representing a single unifying sort of computation. For example, the conversion of one molecule into another by an enzymatic process is a computation as it changes the set of interactions the matter from the original molecule can engage in. Likewise, the movement of a molecule from a cell into the intercellular space -- or vice versa -- also changes the set of interactions that molecule can engage in. We introduce the term multicules to denote the units of this unifying computational representation.

Additionally, for interfacing with experimental research the model must be applicable to systems where data for values of $c$ is readily obtained while data for values of $x$ is more limited, even though the dynamics themselves are in truth defined in terms of $x$. One option is to assume a given estimate of $c$ maps onto an estimated $x$ (and vice versa), and have the model be a hybrid of dynamics over $c$ and $x$. For the purposes of this study, the approach will be to assume a pseudo-invertable\footnote{See Supplemental Information Section 1 for a discussion of the validity of the assumption of pseudo-invertability.} linear map $\mathbf{B}x = c$ which is time-invariant. Incorporating noise will be relevant later, so we will use the notation $\mathbf{\tilde{B}} = \mathbf{B}x + \xi$ with $\xi$ being a vector of zero-mean noise. For the time being, however, we will assume perfect "estimates" and ignore the $\xi$ component of $\mathbf{\tilde{B}}$. Further, we define a vector of inputs to the ecology $u$, and a matrix $\mathbf{Z}$ mapping inputs to multicules. However, it is not in fact required that we have any knowledge of $\mathbf{B}$, $\mathbf{G}$, or any other parameter matrices a priori. This is where the application of reinforcement learning will become central to later sections of the study. For now, we must first define a model of the evolutionary process. The rest of this section will be a discussion of the model and how we can convert it into a Partially Observable Markov Decision Process (POMDP) so as to enable the use of the extensive research in control systems engineering and the theory of artificial intelligence. 

One would ideally like there to be a fairly straightforward means of using existing experimental data to determine an initial model which the RL algorithm will then optimize. In what follows, we derive a model whose complexity can appear unreasonable. Indeed, our learning algorithm would produce a significantly underspecified set of matrices if it alone were used for inferring the model. The added complexity, however, is intentional and strategic for two reasons. First, it allows more direct input of known experimental values, which serve as a Bayesian prior for the model to be learned. This includes data on enzymatic reactions, estimates of the relationship between concentrations of different multicules, population fractions of different cell types, and more. Second, one can use the known values as an approximate ground truth for specifying various parameters, reducing -- potentially quite substantially -- the number of parameters to be derived from the model inferred via Reinforcement Learning. While this second approach may not be considered important in some use cases, it could be useful for many applications. This combination of benefits from the more mechanistic model justifies the complexity, and if a fully mechanistic model is not desired then one can simply use the full mathematical formulation for creating a sort of experiment-driven Bayesian prior over the space of possible models. 

\subsection{Dynamical System Model for Multicule Evolutionary Dynamics}

To begin, we want to ensure the underlying mathematical model is described entirely with respect to multicule "computations" on $x$. As derived in Section 1 of the Supplemental Information, converting the standard EGT model to one based on multicules yields the following model.

\begin{equation}\label{full_cell_model_in_section_the_model}
    d_t c = \mathbf{M}\mathbf{C_B}^T\left(\mathbf{G}x - \vec{\hspace{0.5mm}1} (\mathbf{B}x)^T(\mathbf{G}x)\right) + \tilde{\mathbf{Z}}u
\end{equation}

$\mathbf{C}_B^T$ is a matrix whose diagonals are the entries of $\mathbf{B}x$. In other words, $\mathbf{C}_B$ is defined as

\begin{equation}\label{diag_matrix_form_cells}
    \mathbf{C}_B^T = \sum_{i=1}^{q} \mathbf{E}_i (\mathbf{B} x)^T\vec{e}_i
\end{equation}

Our definition of $\mathbf{B}$ is such that this would be equivalent to the earlier $\mathbf{C}_a$ if our inferences were perfect, but this would require a perfect linear model of all molecular interactions involved -- a fundamental impossibility given the nonlinearity of biological processes. Additionally, values of $c$ -- and thus $\mathbf{C}_a$ -- would be noisy estimates from data.

We note a potentially non-obvious feature of \cref{full_cell_model_in_section_the_model}. We refer to this model as being a quadratic dynamical system, yet it is clear that the term $\vec{1}(\mathbf{B}x)^T(\mathbf{G}x)$ is quadratic but will then be multiplied by $\mathbf{C}_B$ -- which is a function of $x$. However, because of the structure of the problem -- as seen in the fact that we have a scalar value multiplied by $\vec{1}$ so that it will be applied to all vector elements of $d_tc$ -- this is equivalent to defining an additional variable. This would be a book-keeping variable, in some sense, such that we have $\vec{1}(\mathbf{B}x)^T(\mathbf{G}x) = \vec{1}x_{\text{book}}$ separate from the actual $x$ entries. Its dynamics would then be $d_t x_{\text{book}} = (\mathbf{B}x)^T(\mathbf{G}x)$ to preserve system structure. As a result, every component of the dynamical system is either of either linear or quadratic order. The same approach can easily be extended to the full model of \cref{dynamical_system_model} and its equivalent 2nd-order equation.

We assume dynamics to alterations in the ecology can be approximated as some multidimensional impulse at time $t_0$ which decays back to a baseline. $\mathbf{U}$ is a diagonal matrix analogous to $\mathbf{C}_B$, but for the vector of inputs $u$ as

\begin{equation}\label{diag_matrix_form_input}
    \mathbf{U} = \sum_{i=1}^{m} \mathbf{E}_i u^T\vec{e}_i
\end{equation}

Other functional forms are of course possible, and in a wide variety of cases we anticipate that our proofs will apply equally to other models. In this study we will only analyze this particular form of ecological model because the ability to construct high-dimensional sigmoids from it allows the model to describe a broad range of ecological processes. This sigmoid model is introduced below.

\subsection{Non-static Ecology}

Additionally, we would like inputs to be able to modify the ecology of the system and physiology of the organisms within it in ways not directly captured by a mapping from $u$ to $x$. To do this, we define

\begin{equation}\label{g_func_first_pass_not_appendix}
    \mathbf{G}(t) = \mathbf{G}_0 + \mathbf{\Delta}\mathbf{U}\mathbf{V}
\end{equation}

It is important to note that the form of $\mathbf{\Delta} \mathbf{U}\mathbf{V}$ captures modifications to the \textit{ecology} of an evolutionary system which are separate from the direct multicules introduced via an external input into the system. The matrices $\mathbf{\Delta}$ and $\mathbf{V}$ are needed to reshape the dynamics of $\mathbf{U}$, and the three matrices in concert describe the effects of inputs on the ecological state of the model just as $\mathbf{Z}$ describes the inputs' direct mapping to $x$ where relevant.

To capture the decay process post-impulse, we use a diagonal matrix of decay rates $\Xi$ modifying $\mathbf{\Delta} \mathbf{U}\mathbf{V}$ into $\mathbf{\Delta}\tilde{\mathbf{U}}e^{-\Xi t}\mathbf{V}$, and have

\begin{equation}\label{g_dynamics_not_appendix}
    d_t \mathbf{G} = -\mathbf{\Delta}\mathbf{\tilde{U}}\mathbf{\Xi}e^{\mathbf{-\mathbf{\Xi}}t}\mathbf{V}
\end{equation}

One can model a shift from one baseline ecology to another simply by having the midpoint between the two baselines be analogous to one impulse decaying backwards in time towards an original $\mathbf{G}_0$, and a second impulse as meeting in the middle with respect to some new baseline $\mathbf{G}_0'$. 

If we proceed through the matrix multiplications as discussed in Supplemental Information Section 1, we derive the completed model of the evolutionary dynamics with the estimation procedures taken into account. 
\vspace{3mm}

\begin{subequations}\label{dynamical_system_model}
\setlength{\fboxrule}{1pt}
\begin{empheq}{alignat = 1}
  d_t x &= \mathbf{B}^{-1}\mathbf{C}_B\left(\mathbf{M}\mathbf{G}x - \vec{\hspace{0.5mm}1}(x^T\mathbf{B}^T\mathbf{G}x)\right) + \mathbf{Z}u \\
  d_t \mathbf{G} &= -\mathbf{\Delta}\mathbf{\tilde{U}}\mathbf{\Xi}e^{\mathbf{-\mathbf{\Xi}}t}\mathbf{V}
\end{empheq}
\end{subequations}

\vspace{3mm}

Note that $\mathbf{Z}$ is equivalent to $\mathbf{B}^{-1}\tilde{\mathbf{Z}}$, to streamline notation. Entries of $\mathbf{Z}$ serve to map external inputs into the multicules they become within the system, in this study according to the functional form of $u$ as decays from a given input point. In terms of pharmaceuticals, the term including $\mathbf{U}$ can be thought of as the effects of active drug molecules, and $\mathbf{Z}u$ as combining the rates of excretion and the rates of metabolism into pharmacologically inactive compounds. As discussed at the beginning of the
Results section, this complex system of equations enables a Bayesian prior hypothesis over the space of possible models and for the purposes of learning can readily reduce to a simpler model, which -- given sufficient experimental data -- can be re-expanded used as a mechanistic Bayesian posterior distribution over the space of possible models. This simpler model will be summarized using $\mathbf{A}$ to denote the largest matrix resulting from performing all the matrix multiplications in \cref{dynamical_system_model}.

\subsection{POMDP for Multicule Evolutionary Models}

Finally, for the sake of our later analysis, it can be useful to convert the system in \Cref{dynamical_system_model} into a single, equivalent higher-order equation. The derivation is somewhat lengthy, so it can be found in Section 1 of the Supplemental Information. The "bookkeeping variable" mentioned earlier enables us to preserve the standard property of EGT models being quadratic dynamical systems. That property will be important later, as regret bounds for model-predictive reinforcement learning are not known for all but a small number of very specific equations and a much smaller number of model classes -- linear, quadratic, and generalized linear models being the only such general classes for which there are to our knowledge adequate bounds.

To condense our notation for the proceeding sections, we will discretize the 2nd-order dynamical model into a Markov Decision Process (MDP) transfer function. 

\begin{equation}\label{full_model_shorthand_notation}
    \Psi(x'|x,u) = \psi(x,u) + \vec{\xi}(x_{k+1} - x_k)\cdot(\delta t) = \frac{d^2 x}{d t^2}(\delta t)^2 + \vec{\xi}(x_{k+1} - x_k)\cdot(\delta t)
\end{equation}

with $\vec{\xi}$ a zero-mean sub-Gaussian noise vector to account for model error (and absorbing the earlier $\xi$ term), multiplied by $(\delta t)$ so that it is proportional to the time-step length. There is thus a parameter $\sigma_\psi$ describing the maximum variance across entries of this sub-Gaussian noise vector. The noise is sub-Gaussian rather than Gaussian to prevent the mathematical difficulties connected with the MDP RL analysis on an unbounded range of outcomes, and this is also physically correct given conservation of mass. Because we can only observe a subset of the system's components at any given time, this MDP is termed "Partially Observable" and is thus a POMDP. Armed with this single-equation POMDP, we proceed to develop the optimal control results.

\subsection{Parameter Learning}

Summarizing the evolutionary game using the time-evolution operator $\Psi(x(k+1)|x(k),u(k))$ from \cref{full_model_shorthand_notation}, describing how the system changes based on its present state and inputs at some time $k$, we can construct a Bellman function for it. The full discussion can be found in Section 2 of the Supplemental Information, but to summarize we have the following:

\begin{equation}
    V(x(k),u(k)) = L(x(\tau),u(\tau)) + \gamma \int_X dx \Psi(x(k+1)|x(k),u(k)) V(x(k+1),u(k+1)
\end{equation}

Here $V(\cdot,\cdot)$ is the value at a given time based on the state and input, and $L(\cdot,\cdot)$ is the reward associated with the final state which would result at time $\tau$ following a given input sequence $\{u(k), u(k+1), ..., u(\tau)\}$ given the current state $x(k)$.

The remaining key component of our derivations is the learning of parameter matrices themselves. Beginning with some set of matrices inferred from experiments, the expectation values and uncertainty in those inferences define respective prior distributions over the parameter matrices. Here we are agnostic as to the method used to generate these prior distributions.

One will inevitably need to select a subset of variables to create a Reduced-Order Model (ROM) for the POMDP $\Psi$, given the enormous dimensionality of the ideal approximation: a Full-Order Model incorporating all multicules involved in the system. There will be a lower bound on the error resulting from the difference between the ROM and the FOM, expressed as the residual, which is the difference between the state resulting from a ROM and that which would be seen for the same variable subset in an FOM. We denote the magnitude of this error as $\varepsilon_O$. A number of authors have described error bounds for various types of dynamical systems, and this continues to be an area of active research.\cite{Benner_2015_projection_based_model_reduction_dynamical_systems,Feng_2021_error_estimation_reduced_order_modeling,dogancic_2022_model_reduction_error_estimation_dynamical_systems,Nicholson_2014_online_estimation_nonlinear_dynamic_systems_error,Zhang_2015_output_error_estimation_model_order_reduction,Chellappa_2020_adaptive_construction_error_estimation_reduced_order_nonlinear_dynamics} To our knowledge the best bound on this error for nonlinear dynamical systems is given by Chellappa et al in \cite{chellappa_2024_error_estimation_model_reduction_nonlinear_dynamics}. Their method is rather technical, and translating their error bounds to the context of our problem in detail is beyond the scope of this paper. We note the existence of such methods both to acknowledge an interesting and highly relevant body of literature, and to indicate the direction one could pursue in proving detailed bounds on $\varepsilon_O$. In addition, there is some fundamental error limit $\varepsilon_{\text{min}}$ below which a quadratic model for a given biological system could not be further optimized, regardless of whether or not it were reduced-order. Thus the problem we analyze here is more precisely one of reinforcement learning to minimize any error beyond the minimum set by those two limits, and optimal treatment selection given such an optimized model.

The error bounds for reinforcement learning of POMDPs proven across \cite{russo_van_roy_posterior_sampling,osband__russo_van_roy_posterior_sampling_reinforcement_learning,osband_van_roy_eluder_dim_model_based_reinforcement_learning} are both state of the art\footnote{Per our literature review and personal communication with B. Van Roy and M. Kochenderfer.} and highly general compared with most results in the field, which makes them very suitable for this problem. A lengthier discussion of these bounds in our case can be found in Supplemental Information Section 4, and relate specifically to the algorithm Posterior Sampling Reinforcement Learning (PSRL).\cite{osband__russo_van_roy_posterior_sampling_reinforcement_learning} In summary, we use their results to derive regret for using the PSRL algorithm to both learn the parameter matrices and select inputs to optimize trajectories through evolutionary space.

Because $\varepsilon_O$ is separate from the $\xi$ term which we would like to minimize -- the goal of using RL to improve our model -- there will be a baseline accumulation of error in addition to $\varepsilon_k$, the error from the estimation of the integral portion of the Bellman operator. The derivation of the function describing $\varepsilon_k$ is discussed in Section 3 of the Supplemental Information. They will be bounded by the sum of their absolute values, so we know that the error from our approximations (as opposed to the error from sub-optimal parameters, which PSRL is used to minimize) will be bounded as $\varepsilon \leq |\varepsilon_k|+|\varepsilon_O| + |\varepsilon_\text{min}|$, as this would be the case if there were only constructive interference among the error values. 

We use $\{\epsilon\}$ to denote the triplet of upper bounds on the errors. First $\varepsilon$, the upper bound on unavoidable error. Second and third are $\epsilon_R$ and $\epsilon_\psi$, the errors between the inferred and actual functions for each of reward and transition process. $\epsilon_R$ corresponds to learning the $\vec{\omega}$ for the reward function if it is not fully specified in advance, and likewise $\epsilon_\psi$ for learning the model $\Psi$.

\begin{equation}
\begin{aligned}
    \mathbb{E}[\tilde{\mathbf{reg}}](T, \Psi, R, \{\epsilon\}) = \tilde{\mathcal{O}}\Bigg(&\sqrt{T}\Bigg[\sigma_R (m+n) \sqrt{\log\left[\frac{\|\vec{\omega}\|_2 \|\{x,u\}\|_{\max}}{\epsilon_R}\right]} \\
    & \hspace{2mm} + \mathbb{E}[\mathcal{K}]\sigma_\psi n \sqrt{ \log\left[\frac{n\|\mathbf{A}\|_1\|\mathbf{A}\|_{\infty}\|x\|_{\max} }{\epsilon_\psi}\right]}\Bigg] \Bigg)
\end{aligned}   
\end{equation}

$\mathbf{A}$ is a combined model parameter matrix being learned,\footnote{See Supplemental Information Section 1 for a discussion of how the linear quadratic model of $\Psi$ can be converted into a single quadratic function.} and $\tilde{\mathbf{reg}}$ is the regret over all learning episodes. Because there are multiple ways of formulating the problem without altering these bounds, as discussed in Supplemental Information Section 1, we do not specify $\mathbf{A}$ in terms of $\mathbf{\Delta}$, $\mathbf{V}$, $\mathbf{\Xi}$, $\mathbf{B}$, $\mathbf{G}$. In each case, it will still hold that $\mathbf{A}\in \mathbb{R}^{\mathcal{O}(n)\times \mathcal{O}(n)}$, so it does not impact our analysis at this level of granularity.  Here $\|\mathbf{A}\|_1$ and $\|\mathbf{A}\|_{\infty}$ are respectively the maximum column-wise and row-wise sums in the matrix, $\|\mathbf{A}\|_2 \leq \sqrt{\|\mathbf{A}\|_1 \|\mathbf{A}\|_\infty}$ is the induced $\mathcal{L}_2$ norm of the matrix, equivalently its largest singular value, and $n$ is once again the dimension of the multicule vector.

One may notice that despite our discussion of $\varepsilon$ and its components $\varepsilon_0$, $\varepsilon_{\text{min}}$, and $\varepsilon_k$, the variable $\varepsilon$ does not appear in the final bound. This is because the expectation of its contribution to the regret is in fact $0$, with small variance on the scale of the state vector lengths we anticipate this algorithm being applied to. A discussion of this and proof of the distributional properties is included in Supplemental Information Section 5.

The sparseness of $\mathbf{G}$, $\mathbf{U}$, and others due to the rather tight bounds which biology itself imposes upon the matrices combine to further improve the bound beyond what we show in an asymptotic bound. That is, multicules are in general only directly convertible into a small number of other multicules, and the impacts of other processes not directly modeled can reasonably be assumed to be distributed in a Gaussian fashion by application of the Central Limit Theorem.  Additionally, it is reasonable to assume that most aspects of physiology will not change drastically over time on average. As a consequence of these factors, for most $\mathbf{A}$ of interest in this problem, $\|\mathbf{A}\|_1$ and $\|\mathbf{A}\|_{\infty}$ will likely be $<<n\delta t$. 

The portion of total regret accessible to learning grows as the $\sqrt{T}$, but this sublinear increase occurs across linear time and so the regret per unit time decays as $\frac{1}{\sqrt{T}}$, meaning that our learning converges. While the unlearnable component of regret grows linearly, it is zero mean and has a variance which converges to zero for large $n$, yielding an effective overall bound.


\subsection{Computational Complexity}

We show in Section 1 of the Supplemental Information that the time complexity for model evaluation is $\mathcal{O}(qn^2)$, and the number of evaluations per timestep for a given degree of error in terms of integral quadrature sample count is calculated in Section 3 of the SI to be $\varepsilon_k(s) = \mathcal{O}\left(n^2\exp\left(\frac{-s^2}{n^2}\right)\right)$.

Supposing we want a final error bound of some specific value $\epsilon_f = \inf \{\epsilon_R,\epsilon_\psi\}$, each round defining a new $k_0$ and requiring assessment across the $\tau-k_0$ planning range, and require a quadrature sample count for each time-step sufficient to reach $\epsilon_f$. We compute the sample number requirement per time step by inverting the asymptotic bound

\begin{equation*}
    s = \mathcal{O}\left(n\sqrt{\ln\left(\frac{n}{\epsilon_f}\right)}\right)
\end{equation*}

This gives samples per timestep analyzed, and given the $\mathcal{O}(s^3)$ complexity of FWLSBQ the total complexity for error $\epsilon_f$ is

\begin{equation*}
    \mathcal{O} = \left(n^6q^3 \cdot \hspace{1mm }^{3}\hspace{-0.5mm}\sqrt{\ln\left(\frac{n}{\epsilon_f}\right)}\right)
\end{equation*}

This bound at first appears concerning given the $n^6$ term, but because of the practical limitations of experiments we expect $n$ to be on the order of hundreds at most. Additionally, the exponential convergence of FWLSBQ, and the physical limits on a useful $\epsilon_f$ imposed by the inevitable level of noise involved in any biological system further diminish the likely computational demands in practice. We know that for each measurement of the system (occurring at intervals of $\text{\underline{t}}$) we evaluate over the planning range of $\tau$ future timesteps. So the asymptotic complexity is then the product of the total operational time $T$, the planning range length $\tau$, and the quadrature complexity per timestep in planning. Thus the method as a whole has time complexity

\begin{equation}
    \mathcal{O}\left(T\frac{\tau}{\text{\underline{t}}} n^6q^3 \cdot \hspace{1mm }^{3}\hspace{-0.5mm}\sqrt{\ln\left(\frac{n}{\epsilon_f}\right)}\right)
\end{equation}

Given the speed limitations of biological processes, we anticipate the timescales available for computational analysis mean that with appropriate refinement for implementation the bounds should make this approach feasible for near-term application.

\section*{DISCUSSION}

Thus far there has been relatively little work applying contemporary advancements in control theory and related disciplines to questions of evolutionary systems. There is a recent discussion of the use of dynamic programming for the optimization of cancer therapy,\cite{gluzman_2020_dynamic_programming_evolutionary_game_theory} however, we have been unable to find any such work beyond that paper. The massive advances in the mathematical study of control systems and artificial intelligence from recent decades have not yet crossed over into the study of evolutionary dynamics and its application to cancer biology. This represents a fertile ground for the use of new tools, the answering of previously unstudied questions, and the framing entirely new questions. 

In this paper we expand the theoretical methods available for evolutionary dynamics via integrating contemporary research in control theory and the theory of artificial intelligence. Though a discussion of the experimental component of a practical learning procedure is beyond the scope of this study given it will likely be a contextual decision based on the desired application of directed evolution (clinical, industrial, scientific, etc.), our theoretical analysis stands on its own as a first-of-its kind approach for addressing one of the central questions of applied evolutionary theory. The interoperability of Computational duality makes it particularly suitable for the monitoring and course-correction of evolutionary processes, given the multicules corresponding to an intercellular compartment offer a ready readout for the state of the system provided there are adequately accurate inference methods for translating back and forth with experimental data. Cell sorting and whole-cell mass spectrometry likewise offer a means of reading out the state  of intracellular multicules at least below the resolution level of organelles. We expect such inference methods to be very feasible.

We see a number of promising directions for future work, including 1) extending the results to nonlinear transforms $c=\mathbf{B}(x)$ rather than the linear one we assume, 2) analyzing the potential impacts of the difference between the timescale of the respective dynamics of $x$ and $u$, 3) tightening the bounds on quadrature estimate by accounting for empirical features of parameter matrices, and of course 4) applying our approach or variants thereof to experimental problems.

The use of reinforcement learning means that there are minimal requirements of a-priori knowledge of a system. Beyond specifying definitions of multicules and input compounds, we could readily leave all parameter matrices unknown at the start of the problem. This will of course increase the number of trials needed to learn a model when compared with learning given partial prior knowledge of parameters, but the worst-case bounds described here are valid for the cases of any amount of prior knowledge, including the case of no such knowledge.

In summary, this study represents three key advancements. First, a novel mathematical formulation of Computational duality for biophysical dynamics which seamlessly links ecology and evolution into a single unified process -- just as is the reality in nature. Second, Computational duality's universality and ability to more richly interface with the mechanistic properties of biosystems has the potential to strengthen the link between theory and experiment in further novel ways. Third, by establishing for the first time the utility of tools from modern artificial intelligence theory cross-applied to the mathematical theory of evolution and the biophysical questions therein, this offers a new explicit link between theoretical biophysics, ecology, evolutionary theory, and artificial intelligence. All three advancements represent interesting directions for future work of a highly interdisciplinary nature.

\subsection*{Limitations of the study}

Because we sought to provide a model that could interact directly with experiments, certain decisions were made which increase the conceptual and computational complexity of the problem. For example, the mathematical model itself is notably more complex than those typically used in evolutionary dynamics. However, each of the modeling decisions was carefully made so that as much prior experimental data as possible could be incorporated into the model before any training might begin. Our goal was to strike a balance between the capacity for integration with physiological experiments -- where simpler models have often struggled -- and the need to retain a model tractable for algorithmic analysis so that rigorous statements could be made about an interpretable model. Our results on PSRL learning bounds inherently translate to the coarse-grained models typical in evolutionary dynamics, so in a sense that result is incorporated here as the simplest limit case of our work.

Ultimately, this paper seeks to be a first step towards incorporating contemporary reinforcement learning with evolutionary dynamics. Significant additional empirical and theoretical work remains, but this provides a foundation for such investigation.

\newpage

\section*{METHODS}


\subsection*{Problem Statement}
Evolutionary game models are necessarily nonlinear dynamical systems, and in practice at present it is generally only possible to take estimates of their states occasionally due to the need for genetic sequencing or molecular testing of some sort. Thus, any effort to direct evolution will require the application of digital control.\cite{franklin_powell_emami_naeini_control_systems,franklin_powell_workman_digital_control} In reality, $\mathbf{F}$ is a phenomenological representation of some underlying nonlinear system describing the dynamics of biochemistry and signaling processes within and between biological cells and organisms. Recent improvements in biological data-gathering technology \cite{dillies_et_al_high_throughput_rna_seq,weinberg_biology_of_cancer} enable the inference of a linear approximation $\mathbf{G}x$ of an underlying nonlinear system for an $n_0$-dimensional vector $x$ of molecular levels within cells. It should be noted that $\mathbf{G}$ will be a sparse matrix, as any given molecule will typically only interact meaningfully with a small number of other molecules. Exceptions absolutely exist, such as the central importance of glucose\footnote{A sugar which serves as the body's main source of energy system-wide.} and ATP\footnote{A high-energy molecule synthesized from a variety of energy sources, which is used to fuel a wide array of cellular processes.}, but by and large $\mathbf{G}$ will be quite sparse.\footnote{That is, most of the elements of the matrix will be zero.}

The complexity of biological systems presents substantial challenges, but this complexity also produces enormous opportunities if methods can be developed for deliberately and precisely controlling them.\cite{kauffman_origins_of_order,krakauer_worlds_hidden_in_plain_sight,weinberg_biology_of_cancer} Given there are modeling principles which it is believed should transfer over to complex, data-driven, real-world cases, Model Predictive Control appears to be the most viable option for solving this complex MIMO\footnote{Multiple Input, Multiple Output} control problem.\cite{rawlings_model_predictive_control,grune_nonlinear_model_predictive_control}

Assuming such a $\mathbf{G}$ as described above, the question will be as follows. Given the ability to modify some comparatively small subset of individual entries of $\mathbf{G}$ in the form $\mathbf{G}_{ij} \rightarrow \mathbf{\hat{G}}_{ij}$ such that $ \mathbf{G}_{ij} - \mathbf{\hat{G}}_{ij}$ is sufficiently small for a given application,\footnote{We do not precisely define sufficiently small precisely because we anticipate it could vary significantly from application to application.} the goal is to develop algorithms for predicting and controlling the evolutionary system. In particular, to make predictions of changes produced by targeted selections of the indices for nonzero $\mathbf{G}_{ij} - \mathbf{\hat{G}}_{ij}$. 

This yields the equation
\begin{equation}
    d_t c = \mathbf{M}\mathbf{C}_a\left(\mathbf{G}x - \vec{1}c^{\hspace{1mm}T}(\mathbf{G}x)\right)
\end{equation}
where we have in essence simply replaced $\mathbf{F}c$ with $\mathbf{G}x$ as per the discussion above, and $\mathbf{M}$ is a mapping describing the average rate of mutations converting one cell type into another.

In the context of reinforcement learning,  policies are sets of control signals which are input into a system. Given an specific objective and system there will be one or more optimal policies and many sub-optimal ones generally across a spectrum of quality. This objective is termed the "reward" or "value" function, and is defined for a specific problem. Finally, "regret" is used as a metric of the efficacy of some algorithm or approach. Put simply, regret -- typically considered as an expectation value -- is the difference between the "reward" obtained via a policy and the optimal reward that could have been obtained via a perfect policy. 

For the problem of directed evolution, the value function will in general be a function of the various cell types being considered, with more desirable types having increasingly positive (or less negative) contributions to the reward. The vector of these cell types' respective frequencies at time $k$ is represented by $x(k)$. However, experimental supplies are not free, and in a medical context treatments carry with them side effects so an additional term will incorporate their costs. Inputs are represented as $u(k)$.

For mathematical tractability, we use a linear reward function which at time $k$ for state $x$ and input $u$ is

\begin{equation*}
    R(x(k),u(k)) = \vec{\omega}^Tx(k) - \vec{\mu}^Tu(k)
\end{equation*}

Specific $\vec{\omega}$ and $\vec{\mu}$ are determined by the goals in some problem of interest. In this study these goals are set by the humans aiming to control some evolutionary process, but extensions to this model could define it based on an ecology itself to study natural evolution.\footnote{For instance, one could set $u(k)$ to zero, use a different ecology function than we do here, and have $R(x(k))$ be simply maximization over replication rates. In other words, survival of the fittest.} 

There are a few options in terms of how to model a system where data is available in the form of values for $c$, while the dynamics themselves are in truth defined in terms of $x$. One option is to assume a given estimate of $c$ maps onto an estimated $x$ (and vice versa), and have the model be a hybrid of dynamics over $c$ and $x$. For the purposes of this study, the approach will be to assume a pseudo-invertable\footnote{See Supplemental Information Section 1 for a discussion of the validity of the assumption of pseudo-invertability.} linear map $\mathbf{B}x = c$ which is time-invariant. Incorporating noise will be relevant later, so we will use the notation $\mathbf{\tilde{B}} = \mathbf{B}x + \xi$ with $\xi$ being a vector of zero-mean noise. For the time being, however, we will assume perfect "estimates" and ignore the $\xi$ component of $\mathbf{\tilde{B}}$. Further, we define a vector of inputs to the ecology $u$, and a matrix $\mathbf{Z}$ mapping inputs to multicules. 

However, it is not in fact required that we have any knowledge of $\mathbf{B}$, $\mathbf{G}$, or any other parameter matrices a priori. This is where the application of reinforcement learning will become central to later sections of the study. For now, we must first define a model of the evolutionary process. The rest of this section will be a discussion of the model and how we can convert it into a Partially Observable Markov Decision Process (POMDP) so as to enable the use of the extensive research in control systems engineering and the theory of artificial intelligence.

\subsection*{Computational Duality and Multicules}\label{taak_multicules}
As mentioned in the introduction, the complexity of evolutionary processes can be enormous. Tracking the positions of even a negligible fraction of molecules involved in such processes is not only infeasible but almost certainly unproductive. However, there is value in being able to capture key features of spatial dynamics as well as more direct biochemical and evolutionary ones. One simplified model which offers a good balance between the two is to analyze the spatial component of a system as a set of discrete partitions.

We can imagine attaching to any molecule a hypothetical string of bits to serve as an address of sorts. One value for the bit string could say that it is within a biological cell, another that it is outside the cell. If more fidelity is desired, one could denote intranuclear,\footnote{That is, within the nucleus of a cell.} cytosolic,\footnote{Within the region between a cell's nucleus and its outer boundary.} or extrecellular. Alternatively (or going further), one could create separate label string values for any given partition of interest, such as one value for the shared space between a group of cells and then additionally one value each for every type of cell as described in the Introduction.

In this framework there will in general be no fundamental difference between a molecule moving between partitions vs. being chemically transformed \textit{as it undergoes the same motion}. That is to say, a process which moves a molecule $\text{mol}_1$ from partition $\text{Par}_1$ to partition $\text{Par}_2$ is mathematically analogous to another process which took $\text{mol}_1$ and converted it into $\text{mol}_2$ as it is moved from $\text{Par}_1$ to $\text{Par}_2$. These two processes are dual to each other, in the sense in which physicists use the term duality. They are equivalent descriptions of the same underlying process. The fact that this is not the physical reality of the situation should in general not hamper its efficacy as a modeling tool, and from an epistemological perspective that is likely all any duality from physics could claim to be as well.

Using $\text{mol}_i$ to denote molecules and $\text{Par}_j$ to denote partitions,\footnote{With $i$ and $j$ arbitrary indices and not related to any other use of the same index variables elsewhere in the paper}, the equivalence of $\{\text{mol}_1,\text{Par}_1\} \rightarrow \{\text{mol}_1,\text{Par}_2\} \equiv \{\text{mol}_1,\text{Par}_1\} \rightarrow \{\text{mol}_2,\text{Par}_2\}$ requires introducing language for the relationship between $\{\text{mol}_1,\text{Par}_2\}$ and $\{\text{mol}_2,\text{Par}_2\}$. We might term these partition-distinguished molecules particules, but will instead use the term multicules on account of the extensions beyond partitions as additional properties -- and to avoid confusion with the particles of physics. We term the duality itself Computational duality.

Computational duality allows us to capture richer spatial aspects of evolutionary dynamics in the molecular reformulation of EGT discussed above, as we can account for molecule placement simply by creating multiple entries in $x$ which correspond to the same physical molecular structure across as many partitions as relevant for a given problem. For $q$ cell types of interest, one extracellular compartment, and $n_0$ distinct molecular structures of interest, we can have up to $n_{\text{max}}=n_0(q+1)$ multicules. It is a set of $n\leq n_{\text{max}}$ multicules which we consider here. 

Computational duality is a particularly interesting proposition in that is is capable of converting essentially any collection of disparate types of biological processes into a single coherent process type. Abstracting small-molecule structural information and position as part of a collective state can, in principle, be extended to discretizations of other properties such as conformational states of proteins, methylation states of DNA. Versions of this concept appear in systems theory in various contexts, and biophysics allows all properties to be represented as abstract analogs to molecular structures, and all processes as abstract analogs to chemical reactions. The general nature of this view is the reason for the term multicule rather than particule. This combined state space is of course computationally intractably large if one wished to include the entirety of most given biological processes, but reduced state spaces can always be chosen for the variables most relevant to a given scientific question. Perhaps even more interesting is that from a mathematical perspective generalized multicules provide a single, unified framework for all of biology at the level of molecular and biochemical physiology.


\newpage


\section*{RESOURCE AVAILABILITY}


\subsection*{Lead contact}


Requests for further information and resources should be directed to and will be fulfilled by the lead contact, Bryce-Allen Bagley (bbagley@stanford.edu). 

\subsection*{Materials availability}


No experimental materials are associated with this study. 

\subsection*{Data and code availability}


\begin{itemize}
    \item Because this study focused on the mathematical and algorithmic aspects of an applied theory problem in mathematical biology, no experimental data were used or produced.
    \item No code was created for this study.
    \item Likewise, no other resources were generated for this study. The entirety of the work can be found between the paper and the supplemental information.
\end{itemize}

\section*{ACKNOWLEDGMENTS}


This work was funded by a research grant from the Stanford University Physician-Scientist Training Program. The study was further supported by grants from the National Institute of Health (17X074, U54CA261717). We would like to thank Mykel Kochenderfer, Arec Jamgochian, and Benjamin van Roy for guidance in literature review and recommendations of relevant papers in the fields of POMDPs and RL, and the former two for helpful feedback on conceptualization in the early stages of the project. We thank Dena Panovska for her thoughtful insights on linking this paper with experimental methods. Additionally, we thank the late Christopher T. Walsh for valuable discussions of underlying concepts in the early stages of the project and for his mentorship. He will be missed.

\section*{AUTHOR CONTRIBUTIONS}


Conceptualization, B.A.B.; methodology, B.A.B.; investigation, B.A.B., N.K.; writing-–original draft, B.A.B., N.K.; writing-–review \& editing, B.A.B. and C.K.P.; funding acquisition, B.A.B. and C.K.P.; resources, B.A.B. and C.K.P.; supervision, C.K.P.

\section*{DECLARATION OF INTERESTS}


We have no competing interests to declare in relation to this paper.

\section*{SUPPLEMENTAL INFORMATION INDEX}




\begin{description}
  \item 1: Derivation of the Evolutionary Dynamics Model. This section contains the intermediate steps between the description of the problem statement and that of the time-evolution operator seen in the body of the paper. 
  \item 2: Model-Predictive Control of the Evolutionary Game. This section describes the Bellman operator and Bellman function, particularly as relates to this paper.
  \item 3: Reward Function Estimation. This section explains the method for approximating the integral within the Bellman equation, and details the estimation error and computational complexity properties associated with it.
  \item 4: Derivation of Regret Bounds. This section details the calculation of the regret bound for the application of the PSRL reinforcement learning algorithm to the evolutionary game control problem we study in this paper. 
  \item 5: Statistical Properties of Unlearnable Error $\varepsilon$. This section details the derivation and proof of properties of the component of error which represents a minimum below which the learning algorithm could not improve, including proving that it does not impact the bound on the expectation value of regret.
  \item 6: A Sketch of an Experimental Protocol. This section gives a rough conceptual description of what an experimental process using our results would consist of, noting that additional work on the algorithms side--beyond the scope of this paper--is needed first.
\end{description}

\newpage


\bibliography{references}

\bigskip


\newpage

\newpage

\section{Supplemental Information 1: Derivation of the Evolutionary Dynamics Model}\label{appendix_a_model_derivation}

\subsection{System of 1st Order Matrix Differential Equations}\label{model_system_1st_order_odes}

\[
    d_t c_i = c_i \left(\left(\sum_j \mathbf{F}_{ij} c_j \right) - c^{\hspace{1mm}T}\hspace{1mm}\mathbf{F}c\right)
\]

as a description for a given type. Extending to all types present, we have 
\begin{equation}
    d_t c = \mathbf{C}_a \left(\mathbf{F}c - \vec{\hspace{0.5mm}1}c^{\hspace{1mm}T}\hspace{1mm}\mathbf{F}c\right)
\end{equation}

Here $\mathbf{C}_a$ is a diagonal matrix whose entries are the elements of $c$, defined explicitly as
\begin{equation}
    \mathbf{C}_a = \sum_{i=1}^q \mathbf{E}_i c^T\vec{e}_i
\end{equation}

 Where here $\vec{e}_i$ is the vector with all zero entries save for a $1$ at entry $i$, and $\mathbf{E}_i$ is the $q\times q$ matrix with the $q$-th diagonal as 1 and all other entries as 0. We have subscripted this matrix $\mathbf{C}_a$ for convenience and cleaner notation in certain more complex calculations to come.

With the parameter matrices $\mathbf{M}$, $\mathbf{B}$, $\mathbf{C}_B$, and $\mathbf{G}$ defined as in the Results section, we proceed through the derivation of the model as respectively a 1st-order system of equations and a single 2nd-order equation. As a reminder, $\mathbf{M}\in \mathbb{R}^{q\times q}$ , $\mathbf{C}_B \in \mathbb{R}^{q\times q}$ , $\mathbf{B} \in \mathbb{R}^{q\times n}$ , $\mathbf{G}\in \mathbb{R}^{q\times n}$, and $\mathbf{Z}\in\mathbb{R}^{q\times m}$ 

\begin{equation}
    d_t c = \mathbf{M}\mathbf{C_B}^T\left(\mathbf{G}x - \vec{\hspace{0.5mm}1} (\mathbf{B}x)^T(\mathbf{G}x)\right) + \mathbf{Z}u
\end{equation}

\[\label{show_symmetry}
    d_t c = \mathbf{C_B}\left(\mathbf{M}^T\mathbf{G}x - \mathbf{M}^T\vec{\hspace{0.5mm}1} (\mathbf{B}x)^T(\mathbf{G}x)\right) + \mathbf{Z}u
\]

Here we will assume that $\mathbf{M}$ is symmetric, which is in general very close to the empirical truth. Additionally each row of $\mathbf{M}$ sums to $1$, so $\mathbf{M} \vec{\hspace{0.5mm}1} = \vec{\hspace{0.5mm}1}$. Additionally, $\mathbf{C}_B$ is defined as

\begin{equation}\label{diag_matrix_form_cells}
    \mathbf{C}_B = \sum_{i=1}^{q} \mathbf{E}_i ((\mathbf{B} x)^T\vec{e}_i)
\end{equation}

In this study $\mathbf{E}_i$ is defined as the matrix where $E_{jk} = 0$ in all cases except where $j=k=i$, where it is $1$. That is, the matrix of all zeros except for a $1$ at the i-th diagonal entry. Thus, the above is equivalently

\[
    d_t c = \mathbf{C}_B\left(\mathbf{M}\mathbf{G}x - \vec{\hspace{0.5mm}1} (\mathbf{B}x)^T(\mathbf{G}x)\right) + \mathbf{Z}u
\]

as the simplified model for the dynamics of $c$. Completing the conversion by our earlier definition that $c = \mathbf{B}x$ is a time-invariant mapping, if we define $\mathbf{B}^{-1}$ as a Moore-Penrose left pseudoinverse of $\mathbf{B}$, we know that $\mathbf{B}^{-1}d_t c = d_t x$. Of note, computing this pseudoinverse is equivalent to SVD, and thus has a time complexity of $\mathcal{O}(qn^2)$. However, this is a one-time cost. Because we are claiming a left pseudoinverse, $\mathbf{B}$ must have linearly independent columns. Since each column corresponds to one multicule, this is almost certainly true in general because the selected multicules for a model will not include all actual multicules in the system. $\mathbf{B}$ thus encodes interactions with what could be thought of as "implicit" multicules, and with sufficiently precise measurement the columns of $\mathbf{B}$ are very unlikely to be linearly dependent for cell/environment types (elements of $c$) of interest. In other words, the two types would need to be so similar with respect to their actions on molecules as to be difficult to distinguish via experiment, and could thus reasonably be considered one type for practical purposes. Intuitively, many mutations do not modify molecular processes. An alternate way of framing this relies less on how types are distinguished, and instead simply saying that if two multicules behave identically with respect to $\mathbf{B}$ -- equivalent to saying their columns are linearly dependent -- then we will simply approximate them as the same multicule and modify other parameter matrices in the model correspondingly.

Noting that $\mathbf{C}_B\vec{\hspace{0.5mm}1} = c = \mathbf{B}x$ and progressing through the derivation

\[
    d_t x = \mathbf{B}^{-1}\mathbf{C}_B \left(\mathbf{M} - \vec{\hspace{0.5mm}1}(\mathbf{B}x)^T\right)\mathbf{G}x + \mathbf{B}^{-1}\mathbf{Z}u
\]

\begin{equation}\label{x_dynamics}
    d_t x = \mathbf{B}^{-1}\mathbf{C}_B\left(\mathbf{M}\mathbf{G}x - \vec{\hspace{0.5mm}1}(x^T\mathbf{B}^T\mathbf{G}x)\right) + \mathbf{B}^{-1}\mathbf{Z}u
\end{equation}

However, because treatments -- the input signals of interest -- are not maintained at a constant and level dose over time, we must account for the dynamics of $\mathbf{G}$. This is equivalent to the dynamics of a matrix $\mathbf{\Delta}\mathbf{U}\mathbf{V}$ as discussed earlier. Because we want the dynamics of $\mathbf{G}$ to deviate from some baseline in proportion with drug concentrations (themselves approximately exponentially decaying after a dose is administered), the functional form should be roughly

\begin{equation}\label{g_func_first_pass}
    \mathbf{G}(t) = \mathbf{G}_0 + \mathbf{\Delta}\mathbf{U}\mathbf{V}
\end{equation}

With $\mathbf{U}$ the matrix defined as the decay from a baseline $\mathbf{\tilde{U}}$, a matrix of inputs (chemical or otherwise) into the system which is itself defined as

\begin{equation}
    \mathbf{\tilde{U}} = \sum_{i=1}^{m} \mathbf{E}_i u(\tilde{t}\hspace{0.5mm})^T\vec{e}_i
\end{equation}

Here $\mathbf{E}_i$ and $\vec{e}_i$ are used just as with the definition of $\mathbf{C}_B$ in \cref{diag_matrix_form_cells}. To clarify dimensions, $m=m$, $\mathbf{U}\in \mathrm{R}^{m\times m}$, $\mathbf{\Delta}\in \mathrm{R}^{q\times m}$, and  $\mathbf{V}\in \mathrm{R}^{m\times n}$ Defining $u(\tilde{t})$ is the magnitude of the dose applied at the last treatment, which occurred at time $\tilde{t}$. For simplicity of notation, in this derivation $t$ will be shifted such that 

The diagonals will themselves be very sparse, since a patient is unlikely to be treated with more than a very small number of medications at most at any given time. The matrix $\mathbf{\Delta}$ is then the degree to which each drug (signal) would effect the column of $\mathbf{G}$ with which it interferes, depending on which type within $c$ that row of $\mathbf{G}$ corresponds to. It is thus a representation of the susceptibilities to different forms of interference. In situations where more than one drug targets a given column of $\mathbf{G}$, separate $\mathbf{\Delta}$ matrices may be defined and summed. The equation would not be changed in any significant sense by this. While more complex functional forms may describe the amount of drug in a disease microenvironment at a given time (termed "bioavailability"), for the purposes of this analysis the nonzero inputs may be treated as impulses followed by exponential decays. A separate exponential parameter $\Xi_i$ will be associated with each decay. Thus, it's equivalent to write \Cref{g_func_first_pass} as

\begin{equation}\label{g_func}
    \mathbf{G}(t) = \mathbf{G}_0 + \mathbf{\Delta}\mathbf{\tilde{U}}e^{-\mathbf{\Xi}t}\mathbf{V}
\end{equation}

We have chosen to represent $\mathbf{\Xi}$ as a diagonal matrix having positive entries simply for clarity's sake, as multiplying it by $-1$ is a visual reminder that this is a decay process.

\begin{equation}\label{g_dynamics}
    d_t \mathbf{G} = -\mathbf{\Delta}\mathbf{\tilde{U}}\mathbf{\Xi}e^{\mathbf{-\mathbf{\Xi}}t}\mathbf{V}
\end{equation}

These results provide the full model of \cref{dynamical_system_model}.

It is important to note three key aspects of this model which may not be superficially clear. As mentioned in the Results subsection Dynamical System Model for Multicule Evolutionary Dynamics, the $\vec{1}x^T\mathbf{B}^T\mathbf{G}x$ term can be represented by an additional "bookkeeping" variable in the system of equations. This preserves the model's quadratic structure, which is necessary for the proofs of regret bounds. When it comes to the other three aspects just referenced, first, the summation within $\mathbf{C}_B$ means that the first equation of the 2-equation system is a summation over a collection of quadratic models. Second, By concatenating $x$ and $u$ and replacing the relevant matrices with larger combined matrices including the terms corresponding to each original matrix, we can make clearer the quadratic structure. However, this is not at all illustrative, so the model is presented as above (and below, for the single 2nd-order equation). Third and finally, while inferring the matrices $\mathbf{\Delta}$, $\mathbf{V}$, and $\mathbf{\Xi}$ is a time-invariant problem, it can also be defined in terms of another "bookkeeping" variable. Specifically, this second "bookkeeping" variable would be defined the time since last input impulse\footnote{$t-\Bar{t}$, as discussed below in the subsection on System Stability.} and the model would remain quadratic by inferring the 2nd-order Taylor series of $\mathbf{\Delta}\mathbf{U}e^{-\mathbf{\Xi}t}\mathbf{V}$ and $\mathbf{\Delta}\mathbf{\Xi}\mathbf{U}e^{-\mathbf{\Xi}t}\mathbf{V}$ instead of the exact functions -- again using the concatenation procedure noted above.

\subsection{Single 2nd Order Matrix Differential Equation}\label{appendix_a_single_2nd_order_model_equation}

For the purposes of our approach, it is helpful to restate the model as a single, 2nd order equation.

\begin{equation*}
    d_{t^2}^{2} x = d_t\mathbf{B}^{-1}\mathbf{C}_B(\mathbf{M}(\mathbf{G}_0 + \mathbf{\Delta}\mathbf{\tilde{U}}e^{-\mathbf{\Xi}t}\mathbf{V})x) - d_t\mathbf{B}^{-1}\mathbf{C}_B\vec{\hspace{0.5mm}1}(x^T \mathbf{B}^T(\mathbf{G}_0 + \mathbf{\Delta}\mathbf{\tilde{U}}e^{-\mathbf{\Xi}t}\mathbf{V})x) + \mathbf{B}^{-1}\mathbf{Z}d_tu
\end{equation*}

From here we can derive the single-equation form of the model.

\begin{align*}
    d_{t^2}^{2} x = & \mathbf{B}^{-1}\frac{d \mathbf{C}_B}{d_t}\left(\mathbf{M}-\vec{1}x^T\mathbf{B}^T\right)(\mathbf{G}_0 + \mathbf{\Delta}\mathbf{\tilde{U}}e^{-\mathbf{\Xi}t}\mathbf{V})x \\
    & + \mathbf{B}^{-1}\mathbf{C}_B \Bigg[\mathbf{M}\Bigg(\mathbf{G}_0 \frac{d x}{d t} + \mathbf{\Delta}\tilde{\mathbf{U}}e^{-\mathbf{\Xi} t} \left( \mathbf{V}\frac{d x}{d t} - \mathbf{\Xi}\mathbf{V}x\right)\Bigg) \\
    & - \vec{1}x^T\mathbf{B}^T \Bigg(2\mathbf{G}_0 \frac{dx}{d t} + \mathbf{\Delta}\mathbf{\tilde{U}}e^{-\mathbf{\Xi}t}\left(2\mathbf{V} \frac{dx}{d t} - \mathbf{\Xi}\mathbf{V}x   \right)\Bigg)\Bigg] + \mathbf{B}^{-1}\mathbf{Z}d_tu
\end{align*}

If we wish the clean it up further for aesthetic purposes, or for clarity in the sort of computational complexity analysis on which this study focuses, we can ignore certain constant factors to get the cleaner form.

\begin{align*}
    d_{t^2}^{2} x = & \mathbf{B}^{-1}\frac{d \mathbf{C}_B}{d_t}\left(\mathbf{M}-\vec{1}x^T\mathbf{B}^T\right)(\mathbf{G}_0 + \mathbf{\Delta}\mathbf{\tilde{U}}e^{-\mathbf{\Xi}t}\mathbf{V})x \\
    & + \mathbf{B}^{-1}\mathbf{C}_B \Bigg[\left(\mathbf{M}-\vec{1}x^T\mathbf{B}^T\right)\Bigg(\mathbf{G}_0 \frac{d x}{d t} + \mathbf{\Delta}\tilde{\mathbf{U}}e^{-\mathbf{\Xi} t} \left( \mathbf{V}\frac{d x}{d t} - \mathbf{\Xi}\mathbf{V}x\right)\Bigg)\Bigg] + \mathbf{B}^{-1}\mathbf{Z}d_tu
\end{align*}

\subsection{Computational Complexity of Model Evaluation}\label{appendix_a_comp_complexity_evaluating_model}

A quick calculation will show the time complexity of evaluating the model. Here we assume trivial algorithms for matrix multiplication, but of course they can be replaced with more efficient algorithms if the dimensions of $x$ and $c$ are sufficiently large to justify it. Recalling again that $\mathbf{C}_B, \mathbf{M} \in \mathbb{R}^{q\times q}$ , $\mathbf{B}^{-1} \in \mathbb{R}^{n\times q}$ , $\mathbf{G}_0\in \mathbb{R}^{q\times n}$, $\mathbf{\Delta}, \mathbf{Z} \in \mathbb{R}^{q\times m}$, $\tilde{\mathbf{U}}, \mathbf{\Xi} \in \mathbb{R}^{m \times m}$, and $\mathbf{V}\in \mathbb{R}^{m\times n}$ we assess evaluating 

\begin{align*}
    d_{t^2}^{2} x = & \mathbf{B}^{-1}\frac{d \mathbf{C}_B}{d_t}\left(\mathbf{M}-\vec{1}x^T\mathbf{B}^T\right)(\mathbf{G}_0 + \mathbf{\Delta}\mathbf{\tilde{U}}e^{-\mathbf{\Xi}t}\mathbf{V})x \\
    & + \mathbf{B}^{-1}\mathbf{C}_B \Bigg[\left(\mathbf{M}-\vec{1}x^T\mathbf{B}^T\right)\Bigg(\mathbf{G}_0 \frac{d x}{d t} + \mathbf{\Delta}\tilde{\mathbf{U}}e^{-\mathbf{\Xi} t} \left( \mathbf{V}\frac{d x}{d t} - \mathbf{\Xi}\mathbf{V}x\right)\Bigg)\Bigg] + \mathbf{B}^{-1}\mathbf{Z}d_tu
\end{align*}

In terms of computational complexity, the dominant term in the above equation is (separated out) $x^T\mathbf{B}^T \mathbf{\Delta}\tilde{\mathbf{U}}e^{-\mathbf{\Xi}t}\mathbf{\Xi}\mathbf{V}x$ because of the fact that the superficially largest term $\mathbf{B}^{-1}\mathbf{C}_B\vec{1}x^T\mathbf{B}^T \mathbf{\Delta}\tilde{\mathbf{U}}e^{-\mathbf{\Xi}t}\mathbf{\Xi}\mathbf{V}x$ is reduced by being a vector multiplication with $\vec{1}$ multiplied by some scalar value. Evaluating this term has complexity $\mathcal{O}(n^2)$, which is what we would expect for a quadratic model with state vector of length $n$ and input vector of length $m<<n$. In the conservative case where we account for $\mathbf{C}_B$ being a linear sum to provide an exactly quadratic model, we will sum over $q$ terms such that we have evaluation complexity $\mathcal{O}(qn^2)$.

\subsection{System Stability}\label{appendix_a_stability}

In the standard EGT model

\[
    d_t c_i = c_i \left(\left(\sum_j \mathbf{F}_{ij} c_j \right) - c^{\hspace{1mm}T}\hspace{1mm}\mathbf{F}c\right)
\]

represents the dynamics of one type, and  
\begin{equation}
    d_t c = \mathbf{C}_a \left(\mathbf{F}c - \vec{\hspace{0.5mm}1}c^{\hspace{1mm}T}\hspace{1mm}\mathbf{F}c\right)
\end{equation}

for a collection of types. As before,  $\mathbf{C}_a$ is a diagonal matrix whose entries are the elements of $c$, defined explicitly as
\begin{equation}
    \mathbf{C}_a = \sum_{i=1}^q \mathbf{E}_i c^T\vec{e}_i
\end{equation}

 Where here $\vec{e}_i$ is the vector with all zero entries save for a $1$ at entry $i$, and $\mathbf{E}_i$ is the $q\times q$ matrix with the $q$-th diagonal as 1 and all other entries as 0. We have subscripted this matrix $\mathbf{C}_a$ for convenience and cleaner notation in certain more complex calculations to come.

 In this model, $\sum_i c_i = 1$ bounds the state space, which is made valid by the $c_i$ being interpreted as being the fraction of a total population currently comprised of the $i$-th type. Summing over all types thus yields 1. This preserves stability in the deterministic case, but is known (and can readily be shown) to be unstable in the stochastic case in the limit as $t\rightarrow\infty$. Discussions of this can be found in standard textbooks on EGT.\cite{nowak_evo_dynamics,sandholm_population_games_evolutionary_dynamics} It also holds if a Markov matrix $\mathbf{M}$ is included to model mutations between types. As discussed elsewhere, such an $\mathbf{M}$ will be a small deviation from the identity matrix, and can to first order be assumed to take on a form where the off-diagonal entries are some small positive number $\epsilon$ and the diagonal entries $1 - (q-1)\epsilon$. Of note, this is not the $\epsilon$ error term used elsewhere in this study, and in thie paragraph we are merely using the standard notation in analysis for a small value. 

However, even in a deterministic case with an idealized $x$ containing all multicules (and thus molecules) in the system, and a perfect $\mathbf{G}$ and $\mathbf{B}$ to exactly map $x\rightarrow c$, stability would not be preserved. This is because $\mathbf{G}$ and $\mathbf{B}$ are \textit{statistical mappings} -- inferences about the relationship between relevant aspects of genotype, phenotype, and other experimental measures of an evolutionary system. At a physical level, it is also the case that biochemical processes do not preserve the material in the system -- if nothing else in the sense of releasing gasses and producing unusable byproducts. We are thus interested in being able to say some things about the stability of our biophysical cybernetics approach.

\begin{equation}\label{x_dynamics}
    d_t x = \mathbf{B}^{-1}\mathbf{C}_B\left(\mathbf{M}\mathbf{G}x - \vec{\hspace{0.5mm}1}(x^T\mathbf{B}^T\mathbf{G}x)\right) + \mathbf{B}^{-1}\mathbf{Z}u
\end{equation}

Here the issue of instability can be resolved by recognizing it is exactly the sort of imprecision control theory must always account for. Measurements of some components of $x$ are our partial observation just as would be the case in any POMDP, and more broadly most control systems. Thus the control signal $u$ can be used to respond to readouts in the same general fashion as any robust model-predictive control system for any application. This is in some sense inevitable, as there must be external input of material and energy for a system to persist, and similarly there is loss of material as waste. Entropy is a harsh taskmaster.

 However, allowing the inputs to alter the dynamical system itself, in the form of $u$ having effects on $\mathbf{G}$ in addition to the more direct effects on $x$, further complicates the situation.

Our complete model

\begin{align*}
    d_{t^2}^{2} x = & \mathbf{B}^{-1}\frac{d \mathbf{C}_B}{d_t}\left(\mathbf{M}-\vec{1}x^T\mathbf{B}^T\right)(\mathbf{G}_0 + \mathbf{\Delta}\mathbf{\tilde{U}}e^{-\mathbf{\Xi}t}\mathbf{V})x \\
    & + \mathbf{B}^{-1}\mathbf{C}_B \Bigg[\left(\mathbf{M}-\vec{1}x^T\mathbf{B}^T\right)\Bigg(\mathbf{G}_0 \frac{d x}{d t} + \mathbf{\Delta}\tilde{\mathbf{U}}e^{-\mathbf{\Xi} t} \left( \mathbf{V}\frac{d x}{d t} - \mathbf{\Xi}\mathbf{V}x\right)\Bigg)\Bigg] + \mathbf{B}^{-1}\mathbf{Z}d_tu
\end{align*}

Using the previously stated assumption that $u$ -- and thus $\mathbf{U}$ -- are the result of impulse inputs followed by a decay back to some baseline $\mathbf{G}=\mathbf{G}_0$ and $u=0$, we have a clean functional form of 

\begin{equation}
    u_j(t-\Bar{t}) = \tilde{u}_j(\Bar{t})e^{-\Xi_{j,j} (t-\Bar{t})}
\end{equation}

Here $\Bar{t}$ is the time of the last impulse input. Since $\Xi$ is a diagonal matrix, $e^{-\Xi t}$ is also diagonal. Discretizing over time, $\mathbf{\Delta}\mathbf{\tilde{U}} \mathbf{\Xi}e^{\mathbf{-\mathbf{\Xi}}t}\mathbf{V}$ -- and $\mathbf{\Delta}\mathbf{\tilde{U}} e^{\mathbf{-\mathbf{\Xi}}t}\mathbf{V}$ similarly -- would represent a set of matrices, one for each of the $k$ steps of width $\delta t$. One could in principle run PSRL \cite{osband_van_roy_eluder_dim_model_based_reinforcement_learning} with the goal of learning each such matrix and increasing the asymptotic complexity by a factor defined by the number of timesteps beyond which any further decay dynamics is irrelevant. However, they are all the same function of $t-\Bar{t}$, so this would be redundant from a learning perspective. 

This ultimately reduces the dynamic nature of $\mathbf{G}$ to a series of time-dependent functions of $t-\Bar{t}$. While the regret bounds offered by PSRL assume time-invariant matrices, the the simple nature of the functional form we assume makes the mapping to a time-invariant model trivial in the periods between impulse inputs. One can imagine functional forms for $\mathbf{G}(t)$ complex enough to warrant that sort of piece-wise learning, which would scale up the time complexity of the learning process by a factor of the average time span between inputs, as defined according to a given timescale of discretization.

\newpage

\section{Supplemental Information 2: Model Predictive Control of the Evolutionary Game}\label{mpc}

This is a nonlinear dynamical system just as with \cref{default_EGT_equation}, and in the next section we proceed to a discussion of how model predictive control can be used to influence it. Finally, for the sake of our later analysis, it can be useful to convert the system in \Cref{dynamical_system_model} into a single, equivalent higher-order equation. The derivation is somewhat messy, so they can be found in Section 1 of the Supplemental Information as well. Recalling our earlier mention of a "bookkeeping" variable, we note that it is not illustrative in the context of the model's dynamics, but does verify the model's property of being a quadratic dynamical system. That property will be important later, as regret bounds for model-predictive reinforcement learning are not known for all but a small number of very specific equations and a much smaller number of model classes -- linear, quadratic, and generalized linear models being the only such general classes for which there are to our knowledge adequate bounds.

To condense our notation for the proceeding sections, we will discretize the 2nd-order dynamical model into a Markov Decision Process (MDP) transfer function. 

\begin{equation}\label{full_model_shorthand_notation}
    \Psi(x'|x,u) = \psi(x,u) + \vec{\xi}(x_{k+1} - x_k)\cdot(\delta t) = \frac{d^2 x}{d t^2}(\delta t)^2 + \vec{\xi}(x_{k+1} - x_k)\cdot(\delta t)
\end{equation}

with $\vec{\xi}$ a zero-mean sub-Gaussian noise vector to account for model error (and absorbing the earlier $\xi$ term), multiplied by $(\delta t)$ so that it is proportional to the time-step length. There is thus a parameter $\sigma_\psi$ describing the maximum variance across entries of this sub-Gaussian noise vector. The noise is sub-Gaussian rather than Gaussian to prevent the mathematical difficulties connected with the MDP RL analysis on an unbounded range of outcomes, and this is also physically correct given conservation of mass. Because we can only observe a subset of the system's components at any given time, this MDP is termed "Partially Observable" and is thus a POMDP. 

As we will soon show, the use of reinforcement learning means that there are minimal requirements of a-priori knowledge of a system. Beyond specifying definitions of multicules and input compounds, we can leave all parameter matrices unknown at the start of the problem. this will of course increase the number of trials needed to learn a model when compared with learning given partial prior knowledge of parameters, but the worst-case bounds described in the Results subsection on Parameter Learning are valid for the cases of any amount of prior knowledge, including the case of no such knowledge. Armed with this single-equation POMDP, we proceed to develop the optimal control results.

For the finite horizon discrete case, which will be used here for both theoretical and practical reasons, we have the following reward function. With $R(x(t),u(t))$ the instantaneous reward function of a given state $x(t)$ and input $u(t)$ at time $t$, and $L(x(t_f))$ the reward associated with the state at the end of the time horizon considered, the reward of some control sequence $u(t_i \rightarrow t_f)$ is

\begin{equation}
    V\big(t_i,x(t_i),u(t_i\rightarrow t_f) \big) = L(x(t_f)) + \sum_{t_i}^{t_f} dt R\big(x(t),u(t),t\big)
\end{equation}

While the functional form of $L(\cdot,\cdot)$ could be defined differently for various applications, here we assume there are relative values which correspond to each cell type in an ecological setting.\footnote{Bearing in mind that this is simply the most general phrasing. Any evolutionary system from tumors to directed evolution for chemical manufacturing each have their own ecology.} These values form a vector which maps from multicule state to the value function. A similar vector maps input chemicals to their associated costs, and we concatenate these vectors into $\vec{\omega}$. We must also constrain $u(k)$ in order to prevent nonsense solutions, as is it mathematically possible but in practice impossible to over-specify the control of every single molecular detail of a system. However, a conservative scalar weight will serve the same purpose. We define $L(\cdot)$ and $R(\cdot,\cdot)$ with corresponding constraints as:

\begin{equation}
    L(x(t_f)) = \vec{\nu}^T\{x(t_f),0,...,0\}
\end{equation}

\begin{subequations}\label{loss_per_step}
\setlength{\fboxrule}{1pt}
\begin{empheq}{alignat = 1}
  R(x(k),u(k)) & = \vec{\omega}^T \{x(k),u(k)\}\\
  ||u(k)||_2 &\leq \theta
\end{empheq}
\end{subequations}

Here we are using $\{x(k),u(k)\}$ to denote the concatenation of the two vectors, and $\{x(t_f),0,...,0\}$ a vector of equal length with all dimensions normally corresponding to $u(k)$ set to zero. The rewards associated with different cell types are defined by the non-zero entries of $\vec{\nu}$ and $\vec{\omega}$. Using $L_2$ to constrain more intense input concentrations and prevent paths which would require unreasonable control inputs, we threshold with respect to $\theta$. This is convenient for incorporation with the analysis of \cite{osband_van_roy_eluder_dim_model_based_reinforcement_learning}, but in principle could take on a different functional form. The inputs will presumably be subtracted from the reward function simply on the basis that they will not be free in experimental settings.

In practice we anticipate that $\vec{\nu}$ and $\vec{\omega}$ will be the same, and this is assumed in our analysis. $R(\cdot,\cdot)$ will be associated with a single sub-Gaussian variance $\sigma_R$, which will be relevant to the derivation of our regret bounds later. 

Additionally, in the general case $\vec{\omega}$ will not be known exactly in advance, and so a mapping must be learned. To clarify this, one can imagine there being a known $\vec{\omega}'$ which is specified for a given problem, but there is some unknown transform which we can approximate as $\mathbf{\Upsilon}$ such that $\vec{\omega} \approx \mathbf{\Upsilon}\vec{\omega}'$. Of course $\mathbf{\Upsilon}\vec{\omega}'$ is itself a vector, so we use its approximation $\vec{\omega}$. If all entries of $\vec{\omega}$ are considered to be known exactly for a specific case, then the terms corresponding to $R(\cdot,\cdot)$ can be removed from the regret bounds derived later.

Discretizing over time, we would like to frame the problem in the form of a Bellman equation, 

\begin{equation}\label{utility_func_general_form}
    V\big(x(k),u(k)\big) = L\big(x(\tau),u(\tau)\big) + \gamma\int_{X} dx \hspace{1mm} \Psi\big(x(k+1)|x(k),u(k)\big)V\big(x(k+1),u(k+1)\big)
\end{equation}

With $V(\cdot,\cdot)$ the reward of a given state and $\gamma$ a discount factor,\footnote{A scaling parameter which prioritizes events closer in the future over those further on.} our optimal input strategy (in controls and AI terms, the "policy") is given by the following, subject to ceasing to evaluate beyond a chosen time-horizon $\tau$.

\begin{equation}
    u^{\hspace{0.3mm}*} = \arg\max_{u(\cdot)} V\big(x(k_0),u(k_0)\big)
\end{equation}

Which evaluates from some time $k_0$ until the planning horizon $\tau$. Here we do not make any assumptions about the relationship between the interval $\text{\underline{t}}$ between noisy observations of the state and the planning interval $\tau$. This may be disregarded in terms of our proofs of bounds, because observations between $k_0$ and $\tau$ will only improve our knowledge beyond the worst case of only one observation each round at $k_0$ itself.

\newpage

\section{Supplemental Information 3: Reward Function Estimation}\label{frank-wolfe_quadrature}

\subsection{Franke-Wolfe Bayesian Quadrature}

Bayesian Quadrature is a well-established collection of methods for computing numerical solutions to probabilistic integrals.\cite{briol_oates_girolami_osborne_fwbq,kanagawa_hennig_adaptive_bayesian_quadrature,osborne_duvenaud_garnett_rasmussen_roberts_ghahramani_active_learning_evidence_bayes_quadrature}

The Franke-Wolfe algorithm has recently been applied to some problems in statistical learning, such as particle filtering \cite{lacoste-julien_lindsten_bach_sequential_kernel_herding} and Bayesian Quadrature \cite{briol_oates_girolami_osborne_fwbq}. In contrast with the Supplemental Information section Derivation of Regret Bounds Below, we will not discuss the definitions and derivations from \cite{briol_oates_girolami_osborne_fwbq} at any noteworthy length, as transferring their results to this case is somewhat more straightforward. 

In essence, Franke-Wolfe Bayesian Quadrature (FWBQ) and it's more accurate cousin Franke-Wolfe Line Search Bayesian Quadrature (FWLSBQ) use the Franke-Wolfe algorithm to optimize over a convex hull of the posterior distribution volumes to which a given prior could map. This enables a discrete integral of the form $\int_{a\in\mathcal{A}}p(a)f(a)da$ to be solved with accuracy converging faster than other quadrature methods, with respect to number of regions $a_i$ sampled for the discretization. 

\subsection{Calculations}

At each step, we want to solve for

\begin{equation*}
    \int_{x_{k+1}\in X}\Psi\big(x_{k+1}|x_k,u_k\big) V\big(x_{k+1},u_{k+1}\big) = \int_{x_{k+1}\in X} \left(\psi\big(x_k,u_k\big) + \vec{\xi}\cdot (\delta t)\right) V\big(x_{k+1},u_{k+1}\big)
\end{equation*}

But since $\psi\big(x_k,u_k\big)$ and $\delta t$ are constant and the noise is additive,

\begin{equation*}
    V(\psi\big(x_k,u_k\big), u_{k+1}) + \int_{x_{k+1}\in X} \left(\vec{\xi}\cdot (\delta t)\right) V\big(x_{k+1},u_{k+1}\big)
\end{equation*}

is equivalent to separating the integral into deterministic and stochastic components

\begin{equation*}
    V(\psi\big(x_k,u_k\big), u_{k+1}) + (\delta t)\int_{x_{k+1}\in X} \vec{\xi}(x_{k+1} - x_k) \hspace{1mm} V\big(x_{k+1},u_{k+1}\big)
\end{equation*}

As noted earlier, with variance on the sub-Gaussian noise $\sigma_\psi$ in the transition process

\begin{equation}
    \vec{\xi}(x_{k+1} - x_k) = \frac{1}{\sqrt{2\pi\sigma_\psi}}e^{-(x_{k+1}-x_k)/(2\sigma_\psi)}
\end{equation}

we can make use of the distribution's properties to cut down on the number of voxels which must be sampled for a given level of accuracy. $\sigma$ may be small enough that we need only sample a section of the space, accepting that the accuracy bounds of FWLSBQ will be multiplicatively limited by the bounds from cutting off after a certain number of $\sigma$. This is readily accomplished by mandating that all voxels considered must be within some Euclidean distance proportional to $\sigma$. That is, all voxels in $k+1$ for which $\|x_{k+1} - x_k\|_2 \leq \kappa (\delta t)\sqrt{\sigma_\psi}$ with $\kappa$ a tuning parameter and $(\delta t)$ included so the parameter need not scale with the timestep size. This helps to reduce the effective sample count $s$ which is needed, reducing the run-time per integration to be performed.

Because the state space is the unit $\mathcal{O}(n)$-cube, is its diameter just $1$, and with $s$ samples the side-length of each voxel is $\frac{1}{\sqrt[n]{s}}$. Thus, we have a polytope diameter of $D = \frac{n}{\sqrt[n]{s}}$ for the n-volume.

Next, $R$ is the radius of the smallest $n$-sphere which will contain our post-transition state space. Specifically, $R = \frac{\sqrt{s}}{2}$, because we need to account for the full region of the state space across which the system could move from one specific time-step to another. 

FWLSBQ has time complexity bounded by $\mathcal{O}(s^3)$ in the number of points sampled, but the approximation error also converges as

\begin{equation*}
    \text{err}_\text{approx}(s) = \mathcal{O}\left(\frac{n^2}{\sqrt[n]{s}}\exp\left(\frac{-s^2}{n^2/(\sqrt[n]{s})^2}\right)\right)
\end{equation*}

Since $n$ will be on the order of hundreds, $\sqrt[n]{s}\approx 1$ for any computationally feasible value of $s$. So the approximation error bound effectively simplifies to 

\begin{equation*}
    \text{err}_\text{approx}(s) = \mathcal{O}\left(n^2\exp\left(\frac{-s^2}{n^2}\right)\right)
\end{equation*}

While at $\mathcal{O}(s^3)$ -- with $s$ the number of samples used to compute the integral -- it is more computationally expensive than other existing quadrature algorithms, FWLSBQ provides exponentially faster accuracy convergence in the number of samples. For a state space of dimension $n$,\footnote{Recalling that $n$ is the cardinality of $x$.} the convergence of the error between the true integral $T_{\int}$ and the estimate $E_{\int}$ goes as

\begin{equation}\label{frank_wolfe_bayesian_quadrature_convergence}
    \varepsilon_k(s) = \text{err}_{\text{approx}}(s) = \mathcal{O}\left(n^2\exp\left(\frac{-s^2}{n^2}\right)\right)
\end{equation}

Evaluation of the model requires $\mathcal{O}(qn^2)$ in terms of the dimensions of the state, observable, and input vectors, and this is shown briefly in Supplemental Information Section 1's subsection on Computational Complexity of Model Evaluation. This is then the time complexity for each sample in a given timestep's quadrature. It is worth noting, however, that because the parameter matrices should in general be sparse,\footnote{This sparseness is anticipated simply via the fact that most molecules involved in an organism's physiology have substantive interactions with only a small proportion of all other relevant molecules.} this upper bound is a decidedly conservative one.

\newpage

\section{Supplemental Information 4: Derivation of Regret Bounds}\label{appendix_derivation_regret_bounds}

In this section, properties of this problem are derived and/or proven (as the case may be) to prove regret bounds for the use of the Posterior Sampling Reinforcement Learning (PSRL) algorithm described in \cite{strens_bayesian_posterior_reinforcement_learning,osband__russo_van_roy_posterior_sampling_reinforcement_learning,osband_van_roy_eluder_dim_model_based_reinforcement_learning}. We will make particular use of the results from \cite{osband_van_roy_eluder_dim_model_based_reinforcement_learning}, which 

\subsection{Minkowski-Bouligand Dimension and Lipschitz Continuity}

The covering number for balls of radius $\alpha$ is $N_\alpha^{cov} \leq \left(\frac{2k\sqrt{d}}{\alpha}\right)^d$ with maximum norm $k$ and dimension $d$. Since we are interested in the norms of the matrices $\mathbf{B}$, $\mathbf{G}$, $\mathbf{\Xi}$, $\mathbf{\Delta}$, and $\mathbf{V}$ we must instead use a matrix norm. Thankfully, there are equivalence relations which connect most categories of matrix norms, and the induced Euclidean norm of some matrix $\mathbf{A}$ is bounded from above as $\|\mathbf{A}\|_2 \leq \sqrt{\|\mathbf{A}\|_1 \|\mathbf{A}\|_{\infty}}$ by special case of Holder's inequality. For matrix norms, $\|\mathbf{A}\|_1$ and $\|\mathbf{A}\|_{\infty}$ have different definitions than their vector counterparts. The former is the maximum column-wise absolute sum, and the latter the maximum row-wise absolute sum. In other words, the maximum $L_1$ norm over columns and rows, respectively. Given we can reasonably expect $m< q << n$, the largest of our matrices will have $n \cdot q$ terms in the cases of $\mathbf{B}$ and $\mathbf{G}$, these are the dimensions which will dominate asymptotic considerations. In the context of the Minkowski-Bouligand dimension, \cite{dubuc_1989_evaluating_fractal_dimension_minkowski_bouligand,mandelbrot1982fractal_including_minkowski_bouligand,tricot1994curves_fractal_minkowski_bouligand}, a means of computing fractal dimension for the assessment of geometric structures, both have a standard dimension in $\mathbb{R}$ of $nq$ (agnostic to shape), which recalling from the Introduction is equivalent to $nq$. For $\mathbf{\Delta}$ and $\mathbf{V}$ the standard dimension is $nm$, as our input vector $m$ is of length $m$ -- but as mentioned these should not dominate calculations in realistic settings and it is quite simple to substitute their dimensions into subsequent results if so. 

So our covering numbers for a matrix $\mathbf{A}$ will be

\begin{equation*}
    N_\alpha^{cov}(\mathbf{B}) \leq \left(\frac{2\sqrt{\mathcal{I}\mathcal{J}\|\mathbf{A}\|_1\|\mathbf{A}\|_{\infty}}}{\alpha}\right)^{\mathcal{I}\mathcal{J}}
\end{equation*}

with $\mathcal{I}\mathcal{J}$ used to denote the arbitrary dimensions so that calculations need not be repeated with each matrix. For convenience, we define $\tilde{\alpha} = \frac{1}{\alpha}$ The M-B dimension is then defined as 

\begin{equation*}
    \text{dim}_{\text{MB}}(X) = \limsup_{\tilde{\alpha}\rightarrow \infty} \frac{\log N_\alpha^{cov}}{\log(\tilde{\alpha})}
\end{equation*}

So once again using the general $\mathbf{A}$ to represent either matrix of interest the M-B dimension of the matrices here is

\begin{equation*}
    \text{dim}_{\text{MB}}(X) = \limsup_{\tilde{\alpha}\rightarrow \infty} \frac{\mathcal{I}\mathcal{J}\log \left(2\tilde{\alpha}\sqrt{\mathcal{I}\mathcal{J}\|\mathbf{A}\|_1\|\mathbf{A}\|_{\infty} } \right)}{\log(\tilde{\alpha})}
\end{equation*}

 Expanding terms, we have

\begin{equation*}
    \text{dim}_{\text{MB}}(X) = \limsup_{\tilde{\alpha}\rightarrow \infty} \frac{\mathcal{I}\mathcal{J}\left(\log(\tilde{\alpha})+\log \left(2\sqrt{\mathcal{I}\mathcal{J}\|\mathbf{A}\|_1\|\mathbf{A}\|_{\infty} } \right)\right)}{\log(\tilde{\alpha})} = \mathcal{I}\mathcal{J}
\end{equation*}

So as we take the limit, the only term which remains is the product $\mathcal{I}\mathcal{J}$, our M-B dimension. This is not a novel result, but is included for clarity's sake in our usage of the bounding results from \cite{osband_van_roy_eluder_dim_model_based_reinforcement_learning}. 

For the Lipschitz constant $\mathcal{K}$, we need only note that the values of $x$ cannot change faster than the rate of the fastest chemical process whose components are included in $x$. Because in most cases computational complexity will restrict us to some $d$-dimensional subset of the true $d'$-dimensional space of chemicals within the diseased region $x'$, we can define $\mathcal{K}$ as 

\begin{equation}\label{lipschitz_continuity_constant}
    \mathcal{K} = \max_{q\in n} \min_{l\in \mathcal{L}_{qr}} \max_{t\in\mathcal{T}} \frac{d x_l}{dt}
\end{equation}

where $t\in\mathcal{T}$ simply represents all time, $q$ is an index in the state vector $x$ and $l$ is an index of the true $d'$-dimensional vector of chemical species. Put in words, if we have defined some element $\mathbf{G}_{qr}$ such that it describes a multi-step process converting $x_r \rightarrow x_q$ via some intermediate sequence of chemicals with indices $l\in \mathcal{L}_{qr}$, the set of chemicals which exist in the true process of converting between $x_r$ and $x_q$. Essentially, this tells us the rate of change in $x_q$ is bounded from above by the rate of the slowest step in the true chemical process converting $x_r\rightarrow x_q$. This is critical, because chemical processes in living cells can have rates which range over multiple orders of magnitude. If we do not wish to consider intermediate steps, or if a given $x_q$ and $x_r$ pair are both in the vector $x$, then we can simply remove the minimization from the definition where relevant. 

The mathematical definition is less clarifying than a verbal one in terms of the physical meaning. If we define some subset of biochemical processes as being "of interest", $\mathcal{K}$ is essentially the fastest anything "of interest" could ever feasibly change within the volume.\footnote{``Feasibly" is admittedly a somewhat unclear term without a lengthy discussion of bounding extremal cases. Suffice it to say that this restricts our discussion to the set of changes which can occur via biochemical means (interpreted loosely), while excluding various outliers like a physical ablation of a tumor, a sudden mass die-off event, or a meteor strike.} So the above optimization problem yields the Lipschitz constant for the entire model.

\subsection{Eluder Dimension}\label{eluder_dimension_section}

From here, we use the definition of the "eluder dimension"\footnote{Whose form in the vector case was first presented in \cite{osband_van_roy_eluder_dim_model_based_reinforcement_learning}.} for the following analysis. Linear and quadratic matrix equations will be the function classes of interest here, and for the sake of generality we will use $\vec{a}$ and $\mathbf{A}$ to represent arbitrary vectors and matrices.

The eluder dimension of a function class $\mathcal{F}$ is essentially the longest possible sequence of estimated states such that the next state is not revealed to within $\epsilon$ certainty. In the context of this study, $\mathcal{F}$ is equivalent to a combination of 1) the transform between states and observations $ \mathbf{\tilde{B}}$, and 2) the outcome for a given policy, as defined by the mathematical model derived in Supplemental Information Section 1. While our model is quadratic, the eluder dimension for quadratic transition processes was proven in \cite{osband_van_roy_eluder_dim_model_based_reinforcement_learning}.

Put precisely, for a function class $\mathcal{F}$, some value $v\in \mathbf{Q}$ is is $(\mathcal{F},\epsilon)\textit{-dependent}$ on $\{v_1,...,v_k\}\in \mathbf{Q}$ \textit{iff}
\begin{equation*}
    \forall f,\tilde{f} \in \mathcal{F}, \hspace{2mm} \sum_{i=1}^k ||f(v_i)-\tilde{f}(v_i)||_2^2 \leq \epsilon^2 \implies ||f(v_i) - \tilde{f}(v_i)||_2\leq\epsilon
\end{equation*}

As a further explanation, this can be summarized as stating that some sequence cannot be learned to $\leq\epsilon$ confidence within $k$ samples using function class $\mathcal{F}$. The $\epsilon$ \textbf{eluder dimension} of $\mathcal{F}$, or $\text{dim}_\text{E}(\mathbb{E}\left[\mathcal{F}\right],\epsilon)$, is the maximal $k$ from the above definition such that $v$ is not $(\mathcal{F},\epsilon)\textit{-dependent}$ on the sequence of $k$ samples beforehand. Put simply, after $k$ iterations we have learned a member of hypothesis function class $\mathcal{F}$ such that given a known input, the expectation of the output would be known to within L2 norm $\epsilon$ error. If we consider the the number of measurements over which the following condition holds,

\begin{equation*}
    \forall f,f' \in \mathcal{F}, \|f(x)-f'(x) \| \leq \epsilon
\end{equation*}

The eluder dimension is the length of the longest possible sequence of measurements until this condition of $\epsilon$-independence will no longer hold.

For a quadratic matrix function $\vec{a}^T\mathbf{A}\vec{a}$ with $\mathbf{A}\in \mathbb{R}^{\mathcal{I}\times \mathcal{I}}$ and $\forall k, \|\vec{a}(k)\|\leq \|\vec{a}(k)\|_{\max}$, the eluder dimension is 

\begin{equation*}
    \text{dim}_\text{E}(\mathbb{E}\left[\mathcal{F}\right],\epsilon_\mathbf{A}) = \mathcal{O}\left( \mathcal{I}^2\log\left[\frac{\mathcal{I}\|\mathbf{A}\|_1 \|\mathbf{A}\|_{\infty} |\vec{a}|_{\max}}{\epsilon_\mathbf{A}} \right]\right)
\end{equation*}

Substituting in the fact that the largest such $\mathbf{A}$ in our model will be in $\mathbb{R}^{\mathcal{O}(n)\times \mathcal{O}(n)}$ and define functions of $x$, the eluder dimension of the model $\psi$ will be

\begin{equation*}
    \text{dim}_\text{E}(\mathbb{E}\left[\mathcal{F}\right],\epsilon_\mathbf{A}) = \mathcal{O}\left( n^2\log\left[\frac{n\|\mathbf{A}\|_1 \|\mathbf{A}\|_{\infty} \|x\|_{\max}}{\epsilon_\mathbf{A}} \right]\right)
\end{equation*}

Because there are multiple ways of formulating the problem without altering these bounds, as discussed in Section 1 of the Supplemental Information, we do not specify $\mathbf{A}$ in terms of $\mathbf{\Delta}$, $\mathbf{V}$, $\mathbf{\Xi}$, $\mathbf{B}$, $\mathbf{G}$. In each case, it will still hold that $\mathbf{A}\in \mathbb{R}^{\mathcal{O}(n)\times \mathcal{O}(n)}$, so it does not impact our analysis. 

If we have a pre-set linear reward function $R(x(k),u(k))$ it will not factor into the regret bounds. However, we could suppose a situation in which the desired $c$ are not known but there is some known desired output. In this case we would formulate it as a linear function of  $\{x,u\}$ (a concatenation of $x$ and $u$) defined by $\vec{\omega}$ such that 

\begin{equation*}
    R(x(k),u(k)) = \vec{\omega}^T\{x,u\}
\end{equation*}

With $\vec{\omega}$ the function to be learned, the eluder dimension is

\begin{equation*}
    \text{dim}_\text{E}(\mathbb{E}\left[\mathcal{F}\right],\epsilon_R) = \mathcal{O}\left( (n+m)\log\left[\frac{\|\vec{\omega}\|_2 \|\{x,u\}\|_{\max}}{\epsilon_R} \right]\right)
\end{equation*}

\subsection{Regret Bounds}\label{appendix_c_regret_bounds_themselves}

As stated in \cref{utility_func_general_form} and \cref{loss_per_step}, the Bellman equation for our system is 

\begin{equation}
    V(x(k),u(k)) = \mathbb{E}[-\vec{\omega}x(k) - \vec{\mu}u(k)] + \gamma \int_{x'\in X} dx \hspace{1mm} \Psi(x,u)V(x'(k+1),u(k+1))
\end{equation}

Subject to the constraint that $||u(k)||_2 \leq \theta$. The expected regret out to a given time horizon is

\begin{equation}
    \mathbb{E}[\tilde{\mathbf{reg}}] = \sum_{k=k_0}^\tau \left[ V_e(x(k),u(k)) - V_e(x(k),u^{\hspace{0.3mm}*}(k)) \right]
\end{equation}

Which is the expectation value of the difference between the reward of a given policy $u(\cdot)$ versus the reward from an optimal policy $u^{\hspace{0.3mm}*}(\cdot)$. 

Recall that $\psi(x,u)$ is equivalent to $\mathbb{E}[\Psi(x,u)]$ because the noise in $\Psi$ is zero-mean, and so we use the former for simplicity given the expectation value is being studied here.\footnote{See the derivation given in Supplemental Information Section 1, and specifically the shorthand notation defined in \cref{full_model_shorthand_notation}.} $\mathcal{K}$ is the Lipschitz constant for the model, defined in \cref{lipschitz_continuity_constant}. Additionally, $\sigma_\psi$ is the variance of the model's sub-Gaussian noise and $\sigma_R$ is the equivalent for the instantaneous reward function. 

Thus, by the results of\cite{osband_van_roy_eluder_dim_model_based_reinforcement_learning}, for desired expected error bounds of all $\epsilon$ with $T$ the total time elapsed, for $\vec{\omega}$ and the largest parameter matrix (denoted as) $\mathbf{A}$, we obtain a regret bound of

\begin{equation}
\begin{aligned}
    \mathbb{E}[\tilde{\mathbf{reg}}](T, \Psi, R, \{\epsilon\}) = \tilde{\mathcal{O}}\Biggl(\sqrt{T} & \Biggl(\sigma_R(m+n) \sqrt{\log\left[\frac{\|\vec{\omega}\|_2 \|\{x,u\}\|_{\max}}{\epsilon_R}\right]} \\ 
    & + \mathbb{E}[\mathcal{K}]\sigma_\psi n \sqrt{\log\left[\frac{n\|\mathbf{A}\|_1\|\mathbf{A}\|_{\infty}\|x\|_{\max} }{\epsilon_\psi}\right]} \Biggr) \Biggr)
\end{aligned}
\end{equation}

The bounding convention $\tilde{\mathcal{O}}$, has been used instead of the standard $\mathcal{O}$ to denote that we are ignoring two specific terms which are $\mathcal{O}\left( \sqrt{\text{dim}_\text{E}(\mathbb{E}\left[\mathcal{F})\right]\sqrt{\log(T)}}\right)$ with $\mathcal{F} = x^T\mathbf{A}x$ and $\mathcal{F} = R(x(k),u(k))$. 

This gives us our total regret bound for learning model parameters and (if relevant) the reward function using the PSRL algorithm discussed in \cite{strens_bayesian_posterior_reinforcement_learning,osband__russo_van_roy_posterior_sampling_reinforcement_learning,osband_van_roy_eluder_dim_model_based_reinforcement_learning}. Our total regret grows as the $\sqrt{T}$, so the regret per unit time decays as $\frac{1}{\sqrt{T}}$, converging nicely.

\newpage

\section{Supplemental Information 5: Statistical Properties of the Unlearnable Error $\varepsilon$}\label{appendix_d_unlearnable_error}

\subsection{Precise worst-case}

We begin with the assumption that measurement errors are normally distributed and thus appear as Gaussian noise within the transition function. While some portion of the model's error can be reduced through the RL algorithm's improvement process, there remains a certain irreducible error component. This irreducible error has contributions from a number of factors discussed in the main text, resulting in a strict lower bound on the total model error. 

We denote this unlearnable error component as $\vec{\varepsilon}$. For each parameter matrix in the final model equation, there is a corresponding error matrix that defines the unlearnable portion of each entry in that parameter matrix. We aim to define $\vec{\varepsilon}$ in a way that does not explicitly depend on the system state vector $\mathbf{x}$. Using $H$ to denote the combined matrix associated with the linear term and $A$ to denote the combined matrix for the quadratic term, we can describe the irreducible model error as follows:

\begin{equation*}
    \{H + \upsilon, A + \Upsilon\}  \rightarrow \{H\mathbf{x} + \upsilon \mathbf{x} , \mathbf{x}^T A \mathbf{x} + \mathbf{x}^T \Upsilon \mathbf{x}\}
\end{equation*}

Since each $\upsilon$ and $\Upsilon$ characterizes unlearnable contributions to $H$ and $A$ respectively, we represent the unlearnable vector $\vec{\varepsilon}$ as   

\begin{equation*}
    \vec{\varepsilon} = \upsilon\mathbf{x} + \vec{1}\mathbf{x}^T \Upsilon \mathbf{x}
\end{equation*}

By nature of the model, we consider $\mathbf{x}$ in terms of concentrations, satisfying $\|\mathbf{x}\|=1$. This normalization implies a natural upper bound on $\vec{\varepsilon}$, since the worst-case alignment of $\mathbf{x}$ with directions of $\upsilon$ and $\Upsilon$ determines how large $\vec{\varepsilon}$ can become. The extreme case arises precisely if $\mathbf{x}$ aligns with both $\|\upsilon\|_2$ and $\|\Upsilon\|_2$, causing them to reinforce each other completely, meaning we can likewise bound the expected value of $\vec{\varepsilon}$. 

\begin{equation*}
    \vec{\varepsilon} \leq\, \vec{1}\bigl(\|\upsilon\|_2 + \|\Upsilon\|_2\bigr) \quad \text{for }n > 1,
\end{equation*}

\begin{equation*}
    \mathbb{E}[\vec{\varepsilon}] \leq \vec{1}(\frac{1}{n}||\upsilon||_2 + ||\Upsilon||_2)
\end{equation*}

This supports the idea that it is possible to define a useful $\vec{\varepsilon}$ solely in terms of model error and not in any way in terms of $\mathbf{x}$.


\subsubsection{Error Terms in Linear and Quadratic Forms}

First, consider the linear term. Let $H$ be the matrix associated with the linear portion of the model, and let $\vec{\varepsilon}$ be the unlearnable error vector. Then, for a state vector $\mathbf{x}$, 

\begin{equation*}
    H(\mathbf{x}+\vec{\varepsilon}) = H\mathbf{x} + H\vec{\varepsilon}
\end{equation*}

In this expression, $H\vec{\varepsilon}$ represents the unavoidable error contribution within the linear subproblem.

Next, we examine the quadratic term, captured by a matrix $A$. Writing the expansion of $(\mathbf{x}+\vec{\varepsilon})^T A (\mathbf{x}+\vec{\varepsilon})$, we have 

\begin{align*}
    (\mathbf{x}+\vec{\varepsilon})^T A (\mathbf{x}+\vec{\varepsilon}) = & \mathbf{x}^T A \mathbf{x} + \mathbf{x}^T A \vec{\varepsilon} + \vec{\varepsilon}^T A \mathbf{x} + \vec{\varepsilon}^T A \vec{\varepsilon}\\
    = & \mathbf{x}^T A \mathbf{x}  + \mathbf{x}^T (A+A^T) \vec{\varepsilon} + \vec{\varepsilon}^T A \vec{\varepsilon}\\
    \leq & ||\mathbf{x}||^2 ||A||_2 + (||A||_2+||A^T||_2) |\mathbf{x}^T\vec{\varepsilon}| + ||\vec{\varepsilon}||^2||A||_2
\end{align*}

\begin{align*}
    (\mathbf{x}+\vec{\varepsilon})^T A (\mathbf{x}+\vec{\varepsilon}) = & \mathbf{x}^T A \mathbf{x}  + \mathbf{x}^T (A+A^T) \vec{\varepsilon} + \vec{\varepsilon}^T A \vec{\varepsilon}\\
    \leq & ||\mathbf{x}||^2 ||A||_2 \cos{\theta_{x,||A||_2}} + (||A||_2+||A^T||_2) |\mathbf{x}^T\vec{\varepsilon}| + ||\vec{\varepsilon}||^2||A||_2 \cos{\theta_{\varepsilon,||A||_2}}
\end{align*}

Above, we use the fact that each sub-expression can be constrained by norms and alignment factors, and we can equivalently represent alignment factors via cosines. Putting the linear and quadratic components together, and then separating the terms representing contributions from the model errors gives the following:

\begin{align*}
    H(\mathbf{x}+\vec{\varepsilon}) + \vec{1}(\mathbf{x}+\vec{\varepsilon})^T A (\mathbf{x}+\vec{\varepsilon}) = & H\mathbf{x} + H\vec{\varepsilon} + \vec{1}\left(\mathbf{x}^T A \mathbf{x}  + \mathbf{x}^T (A+A^T) \vec{\varepsilon} + \vec{\varepsilon}^T A \vec{\varepsilon}\right)
\end{align*}

\begin{align*}
    H\vec{\varepsilon} + \vec{1} \left(\mathbf{x}^T (A+A^T) \vec{\varepsilon} + \vec{\varepsilon}^T A \vec{\varepsilon}\right) \\
    \leq & H\vec{1}(||\upsilon||_2+||\Upsilon||_2)  + \vec{1} \bigg(\mathbf{x}^T (A+A^T) \vec{1}(||\upsilon||_2+||\Upsilon||_2) \\
    & + \vec{1}^T(||\upsilon||_2+||\Upsilon||_2) A \vec{1}(||\upsilon||_2+||\Upsilon||_2)\bigg)\\
\end{align*}

This bound arises from the fact that each $\vec{\varepsilon}$ term can be aligned with the corresponding largest singular directions, yielding the worst-case additive effect from the irreducible error vectors acting through $H$ and $A$.

\subsubsection{Regret Bound at Time t}
We're not worried about the presence of $\mathbf{x}$, focusing solely on the unavoidable portion of the model error. In particular, at a given time-step $t$, the expected contribution to regret from these unlearnable errors gives the following upper bound:

\begin{align*}
    \mathbb{E}[\text{\textbf{reg}}_{\vec{\varepsilon}}(t)] \leq & \Bigg|\vec{\omega}\Bigg(||H||_2\vec{1}(\frac{1}{n}||\upsilon||_2+||\Upsilon||_2)  + \vec{1} \bigg(\mathbf{x}(t)^T ||(A+A^T)||_2 \vec{1}(\frac{1}{n}||\upsilon||_2+||\Upsilon||_2) \\
    & + \vec{1}^T(\frac{1}{n}||\upsilon||_2+||\Upsilon||_2) ||A||_2 \vec{1}(\frac{1}{n}||\upsilon||_2+||\Upsilon||_2)\bigg)\Bigg)\Bigg|
\end{align*}

Equality occurs if all relevant directions align to maximize each error term. 

For clarity, a more explicit expression that allows for the direct analysis of these angles is the original form:

\begin{equation}
    \vec{\omega}H\vec{\varepsilon} + \vec{\omega}\vec{1}\left( x^T(A+A^T)\vec{\varepsilon} + \vec{\varepsilon}^T A \vec{\varepsilon}\right)
\end{equation}

In the worst-case scenario, all directions are exactly identical, i.e., $\vec{\omega}$ and $H\vec{\varepsilon}$ are in the same direction, $\vec{\varepsilon}$ is in the same direction as the vector for $||H||_2$, $\vec{x}$ is in the same direction as $(A+A^T)\vec{\varepsilon}$, $\vec{\varepsilon}$ is in the same direction as $||A+A^T||_2$, $\vec{\varepsilon}$ is also in the same direction as $(A+A^T)\vec{\varepsilon}$, and $\vec{\varepsilon}$ is in the same direction as the vector for $||A||_2$.

To get a realistic bound, it will be helpful to account for the spectral distributions of $A$ and $H$. For example, the effects of each singular value of $H$ on $\vec{\varepsilon}$ will be rescaled by the cosine of the angle between $\vec{\varepsilon}$ and that singular value's direction. A similar argument applies for $A$, especially when considering $(A + A^T)$, which has its own spectral structure.

\subsection{PDF Calculation}
Let $X$ be a random vector uniformly distributed on the sphere $S^{n-1} \subset \mathbb{R}^n$. Here we have an $(n-1)$-dimensional manifold, which is in $\mathbb{R}^n$ rather than $\mathbb{R}^{n-1}$. A simple example is a circle, which can be parametrized by $\theta$ with polar coordinates, so it is a 1-manifold, yet the points on a circle are given with the 2-d coordinates $(x,y)$. The probability measure is the normalized $(n-1)$-dimensional surface area measure on $S^{n-1}$. This means the entire sphere will have measure 1, and there is a surface area measure $\sigma$, and integrating a function $f$ with respect to $\sigma$ gives the notion of surface area $\times$ some function value. To make $\sigma$ a probability measure, it is necessary that $P(S^{n-1}) = 1.$

\begin{equation*}
    P(A) = \frac{\sigma(A)}{\sigma(S^{n-1})}
\end{equation*}

for all measurable sets $A \subseteq S^{n-1}$. Then $P(A)$ is the probability that $X$ lies in $A$.

Without loss of generality, let $v$ be a fixed unit vector in $\mathbb{R}^n$. We can make this assumption since scaling a non-zero reference vector to be unit length does not impact angle. Define the angle $\Theta$ between $X$ and $v$ by 

\begin{equation*}
    \Theta = \arccos(X \cdot v) \text{ and } 0 \le \Theta \le \pi
\end{equation*}

Note that $\Theta$ is taken to be the smaller angle between the two vectors, which justifies the given interval $[0,\pi]$. 

We wish to derive the probability density function of $\Theta$ in terms of the dimension $n$. Equivalently, this is the same as knowing how large a ``slice" of the sphere corresponds to vectors at angle $\theta$ from $v$. This is only because we have $X$ uniformly distributed on the sphere. 

\subsubsection{Geometric Slicing Argument}
A point $X\in S^{n-1}$ is at an angle $\theta$ from $v$ if and only if $X\cdot v = \cos(\theta)$. The set $\{X : X\cdot v = \cos(\theta)\}$ is an $(n-2)$ spherical cap boundary, meaning it is a slice of the sphere of dimension $(n-2)$. The reason this is $(n-2)$-dimensional is because the sphere itself is $(n-1)$ dimensional, and enforcing a linear condition $(X \cdot v = \cos(\theta))$ reduces the dimension by 1. A familiar example of this is the ordinary 2D sphere in three dimensional space, the set of points on $S^2$ satisfying $X\cdot v = \cos(\theta)$ is a circle $S^1$ which is 1-dimensional.

Consider a slice of the sphere corresponding to angles in $[\phi, \phi + d\phi]$. If we fix $\phi$, the intersection $\{X: \Theta\approx \phi\}$ is an $(n-2)$-dimensional sphere of radius $\sin(\phi)$ sitting in a plane orthogonal to $v$. The measure of that slice is 

\begin{equation*}
    d(\text{Area}) \propto(\sin\phi)^{n-2}d\phi
\end{equation*}

It is easy to visualize this on the 3D sphere. Pick a "north pole" vector $v$, fix an angle, the condition where $X \cdot v = \cos(\phi)$ geometrically slices a $(n-2)$-dimensional manifold. If one thinks about it in terms of a right triangle, $\|X\|=1$, $X \cdot v = \|X\|\|v\|\cos(\phi)=\cos(\phi)$, so by the Pythagorean theorem, the distance orthogonal to $v$ becomes $\sqrt{1-\cos^2(\phi)} = \sin(\phi)$. Hence that slice is all of the vectors whose ``height" is $\cos(\phi)$, living in an $(n-2)$-dimensional space orthogonal to $v$, of radius $\sin(\phi)$. 

In order to get the exact constant of proportionality, we impose an integral from $0$ to $\pi$ of that measure which equals the total surface area of $S^{n-1}$.

\begin{equation*}
    \int_{0}^{\pi} c(\sin\phi)^{n-2}d\phi = \text{Area}(S^{n-1}) = \frac{2\pi^{n/2}}{\Gamma(\tfrac{n}{2})}
\end{equation*}

We then use this known trigonometric integral identity to solve for $c$, and let $m = n - 2$.
\begin{equation*}
    \int_{0}^{\pi} (\sin\phi)^{m}d\phi = \frac{\sqrt{\pi}\Gamma(\tfrac{m+1}{2})}{\Gamma(\tfrac{m+2}{2})}, 
\quad m > -1
\end{equation*}

\begin{equation*}
    \int_{0}^{\pi} (\sin\phi)^{n-2}d\phi = \frac{\sqrt{\pi}\Gamma\!\Bigl(\tfrac{n-1}{2}\Bigr)}{\Gamma\!\Bigl(\tfrac{n}{2}\Bigr)}
\end{equation*}

\begin{equation*}
    c\frac{\sqrt{\pi}\Gamma\!\Bigl(\tfrac{n-1}{2}\Bigr)}{\Gamma\Bigl(\tfrac{n}{2}\Bigr)} = \frac{2\pi^{n/2}}{\Gamma(\tfrac{n}{2})} \Rightarrow c = 
\frac{2\pi^{n/2}}{\Gamma(\tfrac{n}{2})} \cdot \frac{\Gamma(\tfrac{n}{2})}{\sqrt{\pi}\Gamma(\tfrac{n-1}{2}\Bigr)}  = \frac{2\pi^{\tfrac{n}{2}}}{\sqrt{\pi}\Gamma\bigl(\tfrac{n-1}{2}\bigr)}
\end{equation*}

This integral is the total surface area. For the pdf of $\Theta$ we must normalize each ring's area by the total surface area using a normalized measure.

\begin{equation*}
    \frac{d(\text{surface area})}{\text{Area}(S^{n-1})} = \frac{c(\sin\phi)^{n-2}}{\text{Area}(S^{n-1})}
\end{equation*}

\begin{equation*}
    p_{\Theta}(\phi)d\phi
= \frac{\text{(surface area of slice between angles }[\phi,\phi+d\phi])}{\text{(total surface area of }S^{n-1}\text{)}}
\end{equation*}

Using this probability density function and our normalized measure, we can obtain the pdf of the angle $\Theta$ between a fixed unit vector $v$ and a random uniformly distributed vector $X$ on $S^{n-1}$. 

\begin{equation*}
    p_{\Theta}(\phi)
= \frac{c(\sin\phi)^{n-2}}{\text{Area}(S^{n-1})}
\end{equation*}

\begin{equation}\label{pdf_derivation}
    p_{\Theta}(\phi)
= \frac{\frac{2 \pi^{n/2}}{\sqrt{\pi}\Gamma(\tfrac{n-1}{2})}
(\sin\phi)^{n-2}}{\frac{2\pi^{n/2}}{\Gamma(\tfrac{n}{2})}}
= \frac{\Gamma(\tfrac{n}{2})}{\sqrt{\pi}\Gamma(\tfrac{n-1}{2})}
(\sin\phi)^{n-2}
\end{equation}

This completes the derivation of the pdf of $\Theta$, the angle between a fixed unit vector and a random, uniformly distributed vector $X$ on $S^{n-1}$.

\subsection{Accounting For The Distribution of Angles}

We first note that $H$ for the linear term, or $A$ for the quadratic term, must have a singular value decomposition given as 

\begin{equation}\label{decomposition}
    H = U_H \Sigma_H V_H^T
\end{equation}

where $\Sigma_H$ is the diagonal matrix of singular values $\{\sigma_i(H)\}$, and $U_H$, $V_H$ are orthonormal bases of $\mathbb{R}^n$ capturing the left and right singular vectors. Acting on an error vector $\vec{\varepsilon}$ then yields $H\vec{\varepsilon}$. In the worst case, we would say that $\vec{\varepsilon}$ aligns perfectly with the top right singular vector $v_1(H)$ of $H$ corresponding to the largest singular value, so that $\|H\vec{\varepsilon}\|$ achieves its maximum $\|H\|\|\vec{\varepsilon}\|$. In general, the true alignment factor is $\|\vec{\varepsilon}\|\cos(\theta_H)$ where $\theta_H$ is the angle between $\vec{\varepsilon}$ and $v_1(H)$. 

An analogous statement applies for the quadratic term $\vec{\varepsilon}^T A\vec{\varepsilon}$. In a worst-case sense, $\|\vec{\varepsilon}^T A\vec{\varepsilon}\|$ can reach $\|A\|\|\vec{\varepsilon}\|^2$ when $\vec{\varepsilon}$ aligns with the top eigenvector of the symmetric part of $A$. More realistically, one obtains a factor $\cos(\theta_A)$ that reduces the magnitude below its maximal value whenever $\vec{\varepsilon}$ is not perfectly aligned with that principal eigenvector.

From \cref{pdf_derivation} where the distribution of the angle $\theta$ is derived, one can see that typical values of $\cos(\theta)$ are small, especially in high dimensions $n$. Indeed, random vectors on the sphere in $\mathbb{R}^n$ are very likely to be nearly orthogonal, which implies a large angle $\theta\approx\tfrac{\pi}{2}$. Consequently, the expectation $\mathbb{E}[\cos(\theta)]$ is significantly less than 1 as $n$ grows. This is discussed further in the next section.

\subsubsection{Deriving a Tighter Bound}
A more realistic scenario than assuming $\|\vec{\varepsilon}\|$ always aligns with the top singular or eigen-directions is to include an angle $\theta_M$ for each matrix-vector pair. For instance, $\|H\vec{\varepsilon}\|$ is more realistically $\leq\|H\|_2 \|\vec{\varepsilon}\|\cos(\theta_H)$, and so on and so forth for the following terms. If we define $\theta_H, \theta_A, \dots$ for each of these matrix-vector pairs, in the worst case scenario each $\theta=0$. More typically, $\mathbb{E}[\cos(\theta_M)] < 1$, where $M$ are matrices $A,H, \dots$. 

\begin{equation*}
    \mathbb{E}[\|H\vec{\varepsilon}\|] \le \|H\|_2\|\vec{\varepsilon}\| \mathbb{E}[\cos(\theta_H)]
\end{equation*}

\begin{equation*}
    \mathbb{E}\bigl[\vec{\varepsilon}^T A\vec{\varepsilon}\bigr] \le
\|A\|_2(\|\vec{\varepsilon}\|^2) \mathbb{E}[\cos(\theta_A)]
\end{equation*}

\begin{equation*}
    \mathbb{E}\bigl[x^T(A+A^T)\vec{\varepsilon}\bigr] \le \|A + A^T\|_2\|\vec{\varepsilon}\|\|\mathbf{x}\|  \mathbb{E}[\cos(\theta_{A,x})]
\end{equation*}

For multiple angles we just multiply the factor corrections, so if $\vec{\varepsilon}$ needs to align with a singular direction of $H$ and simultaneously align with the top eigenvectors of $A$ we then have the factor $\mathbb{E}[\cos(\theta_H)\cos(\theta_A)]$. 

We know $\vec{\varepsilon}$ is uniformly distributed over the unit sphere, and $\vec{\omega}$ is an external vector whose direction might be fixed, so we can't have distributional arguments for angles involving $\vec{\omega}$. The angles $\theta$ in question will then admit the distribution derived in \cref{pdf_derivation}.

\begin{equation*}
    \bigl|\vec{\omega} H \vec{\varepsilon}\bigr|\cos(\theta_{\vec{\omega},H\vec{\varepsilon}})
\end{equation*}

\begin{equation*}
    ||\vec{\omega}||\cos(\theta_{\vec{\omega},\vec{1}})\cdot\Bigl(||\vec{\varepsilon}||\cos(\theta_{x,(A+A^T)\vec{\varepsilon}})\Bigr)
\end{equation*}

\begin{equation*}
    ||\vec{\omega}||\cos(\theta_{\vec{\omega},\vec{1}})\cdot\Bigl(||\vec{\varepsilon}||^2\cos(\theta_{\vec{\varepsilon},A\vec{\varepsilon}})\Bigr).
\end{equation*}

Recall \cref{decomposition}, then for a vector $\vec{\varepsilon}\in \mathbb{R}^n$

\begin{equation*}
    H \vec{\varepsilon} = \sum_{i=1}^n \sigma_i(H)\,\langle \vec{\varepsilon}, w_i(H)\rangle u_i(H)
\end{equation*}

with $w_i(H)$ the columns of $W_H$ and $u_i(H)$ the columns of $U_H$. For an isotropic $\vec{\varepsilon}$ on the sphere, each $\langle \vec{\varepsilon}, w_i(H)\rangle$ behaves like a mean-zero random variable with variance depending on $\|\vec{\varepsilon}\|$. Summing up these contributions provides 

\begin{equation*}
     \|H\vec{\varepsilon}\|^2 = \Bigl\|\sum_{i=1}^n \sigma_i(H) \langle \vec{\varepsilon}, w_i(H)\rangle u_i(H)\Bigr\|^2.
\end{equation*}

$\|H\vec{\varepsilon}\|\le \|H\|\|\vec{\varepsilon}\|$ in the extreme alignment case, and more generally one obtains 

\begin{equation*}
    \|H\vec{\varepsilon}\| \le \|H\|\|\vec{\varepsilon}\| \cos(\theta_H)
\end{equation*}

which reflects a partial rather than full worst-case alignment.



Now, for the $\mathbf{x}^T(A + A^T)\vec{\varepsilon}$ term, $\mathbf{x}$ is unit length so we can skip that factor. We decompose $(A + A^T)$ via SVD or spectral decomposition, and let $\{v_i\}_{i=1}^n$ be singular values of $(A + A^T)$. let $\{u_i\}$ and $\{w_i\}$ be the left and right singular vectors respectively. It follows that 

\begin{equation*}
    (A + A^T)\vec{\varepsilon} = \sum_i v_i \langle \vec{\varepsilon}, w_i\rangle u_i
\end{equation*}

The direction $\vec{\varepsilon}$ in the basis of $\{w_i\}$ can be random if $\vec{\varepsilon}$ is isotropic, so $\langle \vec{\varepsilon}, w_i\rangle$ is a random scalar with zero mean and distribution consistent with uniform direction on the sphere. The sum $\mathbf{x}^T(A + A^T)\vec{\varepsilon}$ becomes

\begin{equation*}
    \mathbf{x}\cdot \sum_i v_i \langle \vec{\varepsilon}, w_i\rangle u_i
= \sum_i v_i \langle \vec{\varepsilon}, w_i\rangle \bigl(\mathbf{x}\cdot u_i\bigr)
\end{equation*}

Each term has its own angle factor, so overall, we get a random sum of the form $\sum_i v_i \varepsilon_i$, if you rename $\varepsilon_i = \langle \vec{\varepsilon}, w_i\rangle$ and some weighting from $\mathbf{x}\cdot u_i$. 

Now we consider the $\vec{\varepsilon}^T A \vec{\varepsilon}$ term. If we consider the singular value decomposition of $A$, 

\begin{equation*}
    A = U_A \Sigma_A  W_A^T
\end{equation*}

By a similar argument to $\|H\vec{\varepsilon}\|$, one has

\begin{equation*}
    \bigl|\vec{\varepsilon}^T A \vec{\varepsilon}\bigr| \le \|A\| \|\vec{\varepsilon}\|^2 \cos(\theta_{\vec{\varepsilon}, A\vec{\varepsilon}})
\end{equation*}

It is important to see that the final expression we get is a sum of sub terms, each with an alignment factor. The naive bound will lump all of these to equal one, but in our case, we have a cosine term. 

We define the worst-case regret expression where all of our cosine factors become 1, giving $$\mathbf{reg}_{\vec{\varepsilon}}^{\text{(worst)}} := |\vec{\omega} H \vec{\varepsilon}| + \|\vec{\omega}\| \Bigl(\|\vec{\varepsilon}\| + \|\vec{\varepsilon}\|^2\Bigr).$$
In the actual scenario, the alignment angles $\theta$ are not always zero, each term is multiplied by a corresponding cosine factor 

\begin{equation*}
    \mathbf{reg}_{\vec{\varepsilon}}^{\text{(actual)}} := |\vec{\omega} H \vec{\varepsilon}| \cos(\theta_{\vec{\omega},H\vec{\varepsilon}}) + \Bigl[\|\vec{\omega}\| \cos(\theta_{\vec{\omega},\vec{1}}) \Bigr] \Bigl(\|\vec{\varepsilon}\| \cos(\theta_{x,(A+A^T)\vec{\varepsilon}}) + \|\vec{\varepsilon}\|^2 \cos(\theta_{\vec{\varepsilon},A\vec{\varepsilon}}) \Bigr)
\end{equation*}

Because $|\vec{\omega}|\cos(\theta_{\vec{\omega},\vec{1}})$ is a value specified by the user, it can be treated as a constant when doing statistical analysis. It thus may be easier to think of it as some $g(\vec{\omega}) = |\vec{\omega}|\cos(\theta_{\vec{\omega},\vec{1}}) = \varphi$. Thus, our regret term can be expressed as:

\begin{equation*}
    \mathbf{reg}_{\vec{\varepsilon}}^{\text{(actual)}} = |\vec{\omega} H \vec{\varepsilon}| \cos(\theta_{\vec{\omega},H\vec{\varepsilon}}) + \varphi\|\vec{\varepsilon}\| \cos(\theta_{x,(A+A^T)\vec{\varepsilon}}) + \varphi\|\vec{\varepsilon}\|^2  \cos(\theta_{\vec{\varepsilon},A\vec{\varepsilon}})
\end{equation*}

One can observe that each of the three remaining cosine factors depends on one or two uniformly distributed directions in $\mathbb{R}^n$. Because these directions are isotropic on the sphere, their distributions are isomorphic under the action of $\mathrm{SO}(n)$ (i.e., any rotation on the $n$-sphere). As a result, one can apply a suitable rotation so that the first vector in each pair is aligned, while the second vector remains isotropic. This transformation causes the probability distributions for all three cosines to be identical. 

Thus, in the specific scenario where we regard these angles as random and identically distributed, it suffices to introduce a single random angle $\theta \in [0, \pi]$. This leads us to the following expression for the distribution of regret from unlearnable error: 

\begin{equation}\label{same_angle_dist}
    \boxed{\mathbf{reg}_{\vec{\varepsilon}}^{\text{(actual)}} \sim \left(|\vec{\omega} H \vec{\varepsilon}| + \varphi\|\vec{\varepsilon}\| + \varphi\|\vec{\varepsilon}\|^2\right) \cos(\theta)}
\end{equation}

Where $\varphi = \|\vec{\omega}\|\cos(\theta_{\vec{\omega},\vec{1}})$ is a constant set by the user, and $\theta$ is a single representative angle capturing the common isotropic distribution of those alignment factors. 

When you compare the actual sum of sub-terms vs worst case sum of sub terms, it is clear to see 

\begin{equation*}
    \frac{\mathbf{reg}_{\vec{\varepsilon}}^\text{(actual)}}{\mathbf{reg}_{\vec{\varepsilon}}^\text{(worst)}}
= \cos(\theta)
\end{equation*}

\begin{equation}
    \mathbf{reg}_{\vec{\varepsilon}}^\text{(actual)} = \mathbf{reg}_{\vec{\varepsilon}}^\text{(worst)} \cos(\theta)
\end{equation}
This follows when we treat the multiple angle factors as identical, which is accurate up to rotation symmetries in high dimensions.
Now, we apply the "blessing of dimensionality", and find the variance of the cosines of $\theta$. 

A random uniform unit vector $U \in S^{n-1}\subset \mathbb{R}^n$ can be generated by sampling a standard normal vector $X\sim \mathcal{N}(0,I_n)$ and then normalizing $U = \frac{X}{\|X\|}.$ Likewise, another independent unit vector $V$ is obtained from an independent normal $Y$. By construction, 
\begin{equation*}
    \langle U, V\rangle = \frac{\langle X, Y\rangle}{\|X\|\cdot \|Y\|}
\end{equation*}

Because $X$ and $Y$ are standard normal and independent, $\langle X,Y\rangle$ has mean zero and variance on the order of $\|X\|\|Y\|$. As $n\to\infty$, we have $\|X\|\approx \sqrt{n}$ almost surely, and the same for $\|Y\|$. It follows that $\langle X,Y\rangle/\sqrt{n}$ becomes approximately a normal random variable with mean 0 and variance 1, by CLT. Hence, 

\begin{equation}
     \langle U,V\rangle \xrightarrow[n\to\infty]{d} \tfrac{Z}{\sqrt{n}}, \quad Z \sim \mathcal{N}(0,1)
\end{equation}

So the dot product is typically of order $1/\sqrt{n}$. For any $\varepsilon > 0$, the probability $\bigl|\langle U,V\rangle\bigr|\ge \varepsilon$ goes to zero as $n\to\infty$. Additionally, recall that $\langle U,V\rangle = \cos(\Theta)$ is the angle between $U$ and $V$, so $\cos(\Theta)\approx 0\implies \Theta\approx \tfrac{\pi}{2}$. Thus in high dimension, random unit vectors tend to be orthogonal with high probability. 

We have found that $\mathbb{E}[\langle U,V\rangle] = 0$, so $\cos(\Theta)$ is centered near $0$. The variance of $\cos(\Theta)$ is on the order of $1/n$, since $\langle U,V\rangle = \cos(\Theta)$, and for large $n$, $\mathrm{Var}\bigl(\langle U,V\rangle\bigr)\approx \mathrm{Var}\Bigl(\tfrac{Z}{\sqrt{n}}\Bigr) = \tfrac{1}{n}$. Because $\Theta \approx \frac{\pi}{2} - \cos(\Theta)$ for small $\cos(\Theta)$, we can do a small-angle approximation. For $\Theta$ near $\tfrac{\pi}{2}$, let $\Theta = \tfrac{\pi}{2}-\delta$. Then $\cos(\Theta) = \sin(\delta)\approx \delta$ (for small  $\delta$) so $\delta\approx \cos(\Theta)\approx \frac{Z}{\sqrt{n}}$. Hence, the angle's deviation $\delta$ from $\tfrac{\pi}{2}$ scales like $O\bigl(\tfrac{1}{\sqrt{n}}\bigr)$, and variance of $\delta$ is then on the order of $\frac{1}{n}$. 

\begin{equation*}
    \Theta \approx \frac{\pi}{2} \pm\ O\Bigl(\frac{1}{\sqrt{n}}\Bigr) \text{ with high probability as } n \text{ increases}.
\end{equation*}

We have thus demonstrated that our actual regret function is much tighter than our worst-case. Define $\mathcal{C} = \cos(\theta)$. With $\mathbb{E}[\mathcal{C}] = 0$ and $\text{Var}(\mathcal{C}) = \frac{1}{n}$, we have shown:

\begin{equation*}
    \mathbb{E}\bigl[\mathbf{reg}_{\vec{\varepsilon}}^{(\text{actual})}\bigr] = \mathbf{reg}_{\vec{\varepsilon}}^{(\text{worst})} \cdot \mathbb{E}\bigl[\mathcal{C}\bigr] \quad\text{with} \quad \mathbb{E}\bigl[\mathcal{C}\bigr]\ll1 \quad\text{with high probability for large }n.
\end{equation*}

\newpage

\section{A Sketch of an Experimental Protocol}\label{appendix_section_experimental_protocol}

While this paper is predominantly focused on the technical, mathematical aspects of the problem, our goal is for this work to have a near-term and direct pathway to practical application. Before this can be done there is a need for additional work extending this paper to address related problems which are beyond the scope of this paper. Such problems include initialization of the model from existing experimental data, determining sample sizes needed for a given experimental application, and other aspects of bridging our results to direct implementation. 

However, even though the precise technical details of our methods and results are likely to be approachable only for those with significant quantitative backgrounds, one of our core goals in the paper is for at least the gestalt of our results to be approachable to those from experimental biomedical research backgrounds. To that end, in this section we will briefly sketch out a rough framework for applying this method to experiments.

In bullet-point format, the overall process would be thus:

\begin{enumerate}\label{experimental_protocol_sketch}
    \item Determine which biomedical system\footnote{The clearest application is to evolutionary processes. However, thanks to our method's suitability for controlling ecological processes, broadly interpreted (e.g. intercellular and intracellular dynamics even in the absence of evolution), other applications may be feasible.} is to be studied.
    \item Describe quantitatively the initial conditions and goal conditions in terms of either biomarkers or relative frequencies of cell types. Additionally determine the time horizon over which this system is meant to be controlled.
    \item Compile existing and available relevant data on biochemical processes, transport processes, and other biophysical and cellular physiology aspects of the system to be studied.
    \item Apply an algorithm (beyond the scope of this paper) to convert this data into an initial model, so that our algorithm can start from an empirically supported model and thus operate more efficiently.
    \item Repeat (a)-(e) below until a desired level of performance in controlling the biomedical system is achieved, or the results are no longer improving at a rate considered worth the cost and time of additional experiments.
    \begin{enumerate}
        \item Run our approach's control algorithm to determine its prediction of inputs. This is determined from the goal conditions and the current model of the system. 
        \item Take its predictions for optimal inputs, and apply them to the system. 
        \item Either at the end of a given experiment, or intermittently throughout an experiment, feed the current output data from the system (as defined by the experimenters in step 1 above) into the reinforcement learning (RL) algorithm used in our method.
        \item The RL algorithm compares the differences between its predictions and the actual experimental outputs.
        \item The RL algorithm updates its model of the system.
    \end{enumerate}
    \item Analyze the final performance, and apply to problems of interest the model produced by our method.
\end{enumerate}

A visual summary is provided in the figure.

\begin{figure}
    \centering
    \includegraphics[width=0.75\linewidth]{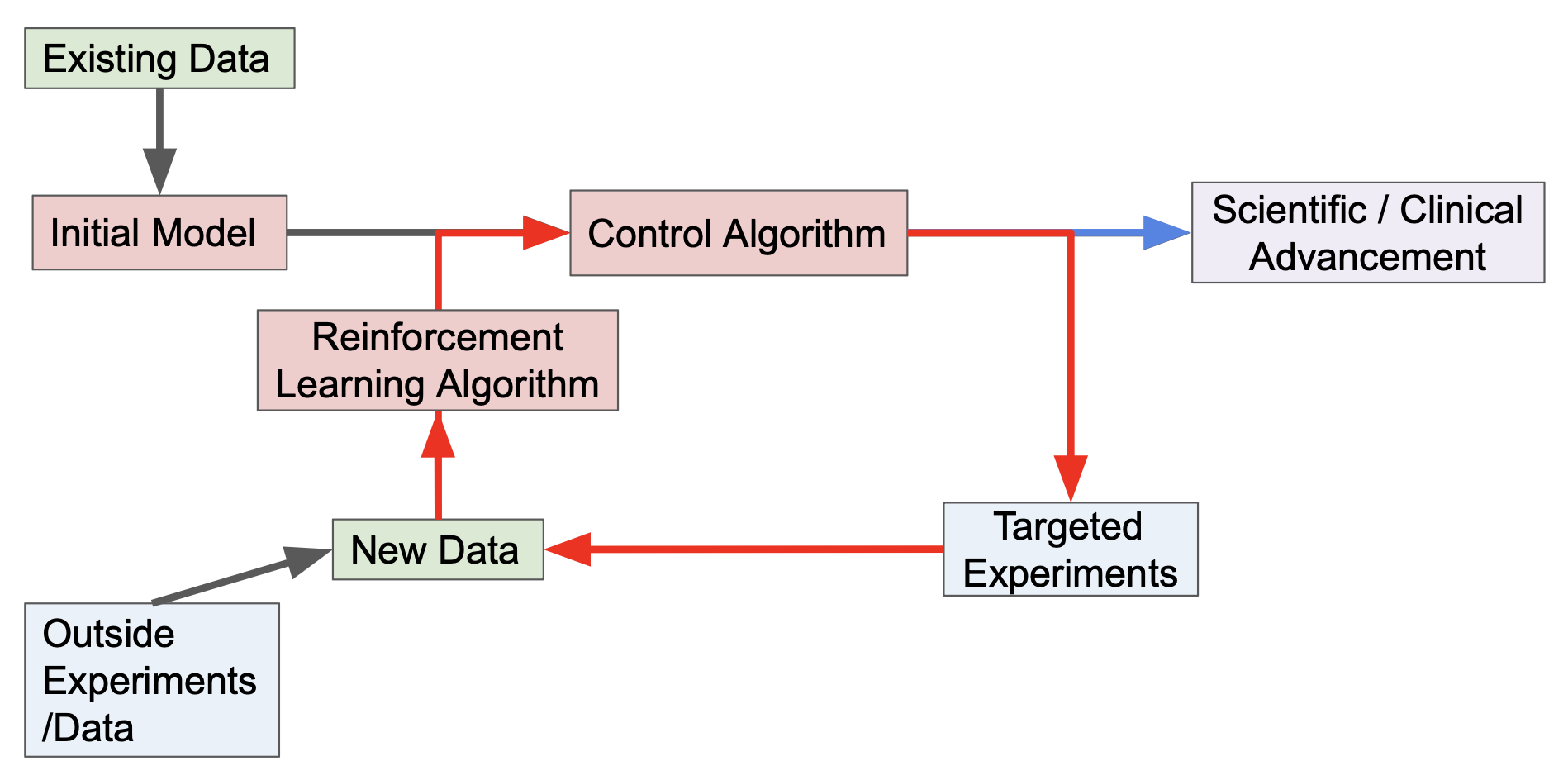}
    \caption{A visual summary of the sketch of an experimental protocol, described above. Red arrows denote the loop described in step 5 of the protocol sketch.}
    \label{fig:experimental_protocol_sketch}
\end{figure}

As noted previously, this is not a thorough protocol at a level used in experimental work. Instead, it is essentially a framework for constructing a thorough protocol, and is meant to help clarify the overall approach of this paper and support collaborations between quantitative researchers and experimental researchers.

\end{document}